\documentclass[aps, prd, twocolumn, preprintnumbers, letterpaper, superscriptaddress, balancelastpage, 10pt]{revtex4-2}
\usepackage{cancel}
\usepackage{lipsum, babel}
\usepackage[utf8]{inputenc}
\usepackage{graphicx}
\usepackage{hyperref}
\usepackage{xcolor}
\usepackage{amsmath, amsfonts, amssymb}
\usepackage{bm, bbm}
\usepackage{enumitem}
\usepackage[caption=false]{subfig}
\usepackage{slashed}
\usepackage{tikz-feynman}
\usepackage[normalem]{ulem}
\newcommand{\beq}{\begin{equation}\begin{aligned}{}}
\newcommand{\eeq}{\end{aligned}\end{equation}}
\newcommand{\beqa}[1]{\begin{equation}\begin{aligned}{#1}}
\newcommand{\eeqa}{\end{aligned}\end{equation}}

\newcommand{\bea}{\begin{eqnarray}{}}
\newcommand{\eea}{\end{eqnarray}}

\definecolor{cadmiumgreen}{rgb}{0.0, 0.42, 0.24}

\begin{document}
%%%%%%%%%%%%%%%%%%%%%%%%%%%%%%%%
%%%%%%%%%%%%%%%%%%%%%%%%%%%%%%%%

% \preprint{
% \begin{flushright}
% \hfill DESY-25-028 
% \end{flushright}
% }
% \preprint{
% \begin{flushright}
% \hfill IFT-UAM/CSIC-25-21
% \end{flushright}
% }
\begin{minipage}{8cm}
\vspace{-1cm}
    \begin{flushright}
IFT-UAM/CSIC-25-21\\
DESY-25-028
\end{flushright}
\end{minipage}

\title{
Inverse bubbles from broken supersymmetry
}
\author{Giulio Barni}
 \email{giulio.barni@ift.csic.es}
\affiliation{Instituto de F\'isica Te\'orica IFT-UAM/CSIC, Cantoblanco, E-28049, Madrid, Spain}
\author{Simone Blasi}
\email{simone.blasi@desy.de}
\affiliation{Deutsches Elektronen-Synchrotron DESY, Notkestr.~85, 22607 Hamburg, Germany}

\author{Miguel Vanvlasselaer}
\email{miguel.vanvlasselaer@vub.be}
\affiliation{Theoretische Natuurkunde and IIHE/ELEM, Vrije Universiteit Brussel,
\& The International Solvay Institutes, Pleinlaan 2, B-1050 Brussels, Belgium}

\begin{abstract}

Building upon the recent findings regarding inverse phase transitions in the early universe, we present the first natural realisation of this phenomenon within a supersymmetry-breaking sector.
We demonstrate that
inverse hydrodynamics, which is characterized by the fluid being sucked by the bubble wall rather than being pushed or dragged, is actually not limited to a phase of (re)heating but can also occur within the standard cooling cosmology. 
Through a numerical analysis of the phase transition, 
we establish a simple and generic criterion to determine its hydrodynamics based on the generalised pseudo-trace. 
Our results provide a proof of principle 
highlighting the need to account for these new fluid solutions when considering cosmological phase transitions and their phenomenological implications.
\end{abstract}

\maketitle

{\it Introduction --} 
Phase transitions (PTs) in the early universe plasma, usually called \emph{cosmological} phase transitions, are fascinating phenomena. First order PTs (FOPTs) proceeding via the nucleation and expansion of bubbles of the true vacuum inside a sea of false vacuum are of particular interest as they can be at the origin of the matter-antimatter asymmetry of the universe (baryogenesis)~\cite{Kuzmin:1985mm, Shaposhnikov:1986jp,Nelson:1991ab,Carena:1996wj,Cline:2017jvp,Long:2017rdo,Bruggisser:2018mrt,Bruggisser:2018mus, Bruggisser:2022rdm,Morrissey:2012db,Azatov:2021irb, Huang:2022vkf, Baldes:2021vyz, Chun:2023ezg}, lead to the production of dark matter~\cite{Falkowski:2012fb, Baldes:2020kam,Hong:2020est, Azatov:2021ifm,Baldes:2021aph, Asadi:2021pwo, Lu:2022paj,Baldes:2022oev, Azatov:2022tii, Baldes:2023fsp,Kierkla:2022odc, Giudice:2024tcp, Azatov:2024crd} and primordial black holes~\cite{Kodama:1982sf,Kawana:2021tde,Jung:2021mku,Gouttenoire:2023naa,Lewicki:2023ioy}, and can be a powerful source of primordial gravitational waves (GWs) as well~\cite{Witten:1984rs,Hogan_GW_1986,Kosowsky:1992vn,Kosowsky:1992rz,Kamionkowski:1993fg}. The broad program to discover and investigate a possible background of GWs by current experiments such as LIGO--Virgo--Kagra~\cite{Romero:2021kby} and Pulsar Timing Arrays~\cite{Bringmann:2023opz}, as well as future detectors such as the LISA~\cite{Caprini:2015zlo} and the Einstein Telescope~\cite{ET:2019dnz},
%with LVK, PTAs, LISA, ET, CE, BBO, Decigo, 
opens the unique opportunity of probing the existence of FOPTs and of new fundamental physics. Indeed, FOPTs appear naturally in a large variety of scenarios beyond the Standard Model (BSM) like composite Higgs~\cite{Pasechnik:2023hwv, Azatov:2020nbe,Frandsen:2023vhu, Reichert:2022naa,Fujikura:2023fbi}, extended Higgs sectors~\cite{Delaunay:2007wb, Kurup:2017dzf, VonHarling:2017yew, Azatov:2019png, Ghosh:2020ipy,Aoki:2021oez,Badziak:2022ltm,Banerjee:2024qiu}, axion models~\cite{DelleRose:2019pgi, VonHarling:2019gme}, dark Yang-Mills sectors~\cite{Halverson:2020xpg,Morgante:2022zvc}, $B-L$ breaking sectors~\cite{Jinno:2016knw, Addazi:2023ftv} and SUSY breaking sectors~\cite{Craig:2009zx, Craig:2020jfv}, and may also be catalysed by impurities in the early universe, see e.g.~\cite{Steinhardt:1981ec,Hosotani:1982ii,PhysRevD.34.1237,Mukaida:2017bgd,Canko:2017ebb,Jinno:2021ury,Balkin:2021zfd,Agrawal:2022hnf,Blasi:2022woz,Agrawal:2023cgp,Jinno:2023vnr,Blasi:2024mtc,Blasi:2023rqi}, as well as occur in the late Universe in the core of neutron stars~\cite{Balkin:2021zfd,Balkin:2021wea,Casalderrey-Solana:2022rrn,Balkin:2023xtr}.

The dynamics of FOPTs involve a non--trivial interplay between the bubble wall and the surrounding plasma, which is pivotal in determining the phenomenology of the PT including the GW emission. The hydrodynamical modes describing the bulk fluid motion in the background of an expanding bubble during a \emph{direct} FOPT have been classified a long time ago~\cite{Laine:1993ey, Kurki-Suonio:1995rrv, Laine:1998jb,Espinosa:2010hh}: they consist of detonations, hybrids and deflagrations. For all these solutions, the fluid is either dragged or pushed (or both) by the bubble wall. The bulk fluid velocity is then always aligned with the wall velocity in the plasma frame\footnote{The plasma frame is defined as the frame in which the centre of the bubble is at rest. }. In the case of the \emph{inverse PTs}, the plasma is instead sucked \emph{inside} the expanding bubble and the fluid flows in the opposite direction of the bubble wall motion~\cite{Barni:2024lkj, Bea:2024bxu} (see also Ref.\,\cite{Cutting:2022zgd} for the case of droplet collapse). These solutions have been so far studied in the context of a (re)heating PT~\cite{Buen-Abad:2023hex,Barni:2024lkj, Bea:2024bxu} where the temperature of the system increases with time, and thus have been associated to superheated bubbles, see also\,\cite{Caprini:2011uz,Dent:2024bhi}.

In this Letter, we show that inverse hydrodynamics is actually not limited to the heating scenario mentioned above, but can instead take place during the standard cooling of the universe as well. 
Remarkably, we find the emergence of this novel hydrodynamics during a seemingly standard, symmetry--breaking phase transition. This takes place within the context of dynamical supersymmetry (SUSY) and $R$--symmetry breaking, which represents the first explicit example for this class of FOPTs.
Our findings extend the current understanding of what types of cosmological phase transitions can actually take place, thus opening up new directions for studying the corresponding GW signatures and other phenomenological aspects.

{\it PTs in a SUSY breaking sector --} Supersymmetry is not a symmetry of the low-energy theory. Therefore, if it is realised at high energy scales, it must be broken by a dedicated SUSY-breaking sector. A broad class of perturbative SUSY-breaking mechanisms can be described within the framework of an effective field theory that encapsulates the dynamics of the so-called \emph{pseudomodulus}. This pseudomodulus corresponds to the scalar component $x$ of the chiral superfield, $X$, which is directly related to SUSY breaking
\bea 
X = \frac{x}{\sqrt{2}}e^{2i a/f_a} + \sqrt{2} \theta \tilde G + \theta^2 F \,, 
\eea 
where we have used the standard superspace notation.  In our study, as a minimal benchmark model, we focus on the  O’Raifeartaigh model~\cite{ORaifeartaigh:1975nky}. In addition to the \emph{pseudo-modulus}, the SUSY breaking sector contains four chiral superfields $\phi_1, \tilde \phi_1, \phi_2, \tilde \phi_2$. The superpotential takes the form
\bea 
\label{eq:W}
W  = -F X + \lambda X \phi_1 \tilde \phi_2 + m(\phi_1 \tilde \phi_1 + \phi_2 \tilde \phi_2) \, .
\eea 
The model preserves a global $U(1)$ $R-$symmetry, which typically accompanies dynamical SUSY breaking~\cite{Nelson:1993nf,Intriligator:2007py}, and under which $X$ has charge two, $R[X] = 2$. The vacuum expectation value (vev) of \( x \) is the order parameter for spontaneous  \( R \)-symmetry breaking, while the additional scalar fields from $\phi_i$ and $\tilde \phi_i$ will have vanishing vev in all phases.
The tree--level vacuum energy is \( V^{\rm min}_{\rm tree} = |F|^2 \) with $x$ being a flat direction, indicating that supersymmetry is broken irrespectively of \( \langle x \rangle \) while \( R \)-symmetry is preserved only at the origin, $\langle x \rangle = 0$. 

The pseudomodulus flat direction is however lifted at the loop level, and the shape of the potential for $x$ is controlled by the mass spectrum of the theory.
One can see that this includes massive particles from the $\phi_i$ and $\tilde \phi_i$ superfields, with the scalar eigenstates split in pairs around the fermion ones, as well as massless fields from the superfield $X$ corresponding to the pseudomodulus, $x$, the $R$--axion, $a$, and the goldstino, $\tilde G$. At one--loop, the potential for $x$ acquires a global minimum at the origin, while remaining remarkably flat at large field values as a reflection of the underlying SUSY.

Finite--temperature effects, on the other hand, break SUSY explicitly and have a strong impact on the pseudomodulus effective potential. The typical thermal history of the minimal O'Raifeartaigh model considered here is then as follows~\cite{Craig:2020jfv,Craig:2006kx, Katz:2009gh}: at very high temperatures, $T \gtrsim \sqrt{F}$, the system has a single vacuum state, $\langle x\rangle =0$, and $R$--symmetry is preserved. At lower temperatures, a new local minimum of the effective potential appears at relatively large field values, $\langle x \rangle/\sqrt{F} \gg 1$, which becomes the true vacuum of the theory below a certain critical temperature, $T_c$. This vacuum with broken $R$-symmetry will however become metastable and eventually disappear at even lower temperatures, given that the only minimum at zero temperature is at $\langle x \rangle = 0$. 

Overall, the system undergoes two phase transitions, namely \textit{(1)} the breaking of the $R$--symmetry at high temperatures and \textit{(2)} its restoration at low temperatures, which turn out to be first order and governed by a thermal barrier. More details on the standard derivation of the effective potential for $x$ and the associated thermal history can be found in App.\,\ref{app:effect_pot} and references therein\,\cite{PhysRevD.7.1888,Quiros:1999jp,Curtin:2016urg}, as well as in Ref.\,\cite{Craig:2020jfv}.

In this paper, we will focus on the first transition that will take place in the expanding universe, namely the $R$--symmetry breaking FOPT: $\langle x=0 \rangle \rightarrow \langle x \neq 0\rangle$. As it turns out, this FOPT can actually proceed according to either the direct or the inverse hydrodynamics (the latter presented in Ref.~\cite{Barni:2024lkj, Bea:2024bxu}) depending on the microscopic coupling constant $\lambda$ entering the superpotential in Eq.\,\eqref{eq:W}, while the second $R$--symmetry restoring FOPT will always be direct. 
 
\medskip 

{\it Thermodynamics and hydrodynamics of $R$--symmetry breaking --} In the early universe, FOPTs can be modelled as the interplay between a scalar field \(\phi\), whose vacuum expectation value represents the order parameter of the transition, and the surrounding plasma which is often well described by a relativistic fluid.  The energy--momentum tensor of the system consists then of those two contributions, $T^{\mu\nu} = T^{\mu\nu}_{\text{fluid}} + T^{\mu\nu}_{\phi}$, with
\begin{subequations}
\begin{align}
T^{\mu\nu}_{\phi} &= \partial^\mu \phi \partial^\nu \phi - g^{\mu\nu} \left( \frac{1}{2} (\partial \phi)^2 - V(\phi) \right)\,,
\\
T^{\mu\nu}_{\text{fluid}} &= (e + p) u^\mu u^\nu - p g^{\mu\nu}\,,
\end{align}
\end{subequations}
where \(u^\mu\) is the four-velocity of the fluid, \(e\) is the energy density, \(p\) is the pressure and \(V(\phi)\) is the scalar potential. The pressure is related to the free energy as $p = - \mathcal{F}$, while the energy and enthalpy density are given by 
\bea 
e = T \frac{dp}{dT} - p \, , \qquad w = e + p  = T \frac{dp}{dT} \, . 
\eea 

In any particle physics model that can be solved (even if only approximately, \textit{e.g.} in a loop expansion), the free energy $\mathcal{F}$ can be obtained directly from the effective potential at finite temperature, $V_0+ V_T \equiv \mathcal{F}$. Consequently, the knowledge of the free energy of a given theory allows us to compute all the thermodynamic quantities of interest without introducing a simplified Equation of State (EoS) for the fluid, such as for instance the bag EoS and its generalizations.

The conservation of the energy--momentum tensor across the phase boundary, $\nabla_\mu T^{\mu\nu} = 0$, gives the following relations between the velocities,  the energies and the pressures\,\footnote{We remind that the velocities $v_\pm$ have to be understood in the \emph{front frame}, where the bubble wall is at rest.}
\bea 
\label{eq:rel_velocities}
v_+v_- = \frac{p_+- p_-}{e_+-e_-}\ ,\qquad \frac{v_+}{v_-} = \frac{e_-+p_+}{e_++p_-}\, ,
\eea 
where the subscript ``$\pm$'' denotes quantities in front of/behind the phase boundary, so that for instance ``$-$'' always represents the interior of the bubble. 

We defined inverse PTs as transitions displaying negative bulk velocities in the plasma frame: rather than being pushed outward, the surrounding plasma is drawn inward, effectively being sucked into the expanding bubble. 
Let us now provide a sharper characterization, or criterion, of inverse hydrodynamics which extends the intuitive one put forward in Ref.~\cite{Barni:2024lkj},
according to which inverse PTs are found when the transition proceeds against the vacuum energy (namely, the $T=0$ effective potential for the order parameter). We find that a fully general characterization of inverse hydrodynamics can be obtained by defining a generalised pseudo-trace, $\alpha_\vartheta$, which indicates the strength of the phase transition and extends the definition within the bag EoS adopted in ~\cite{Barni:2024lkj} as well as the pseudo-trace, $\alpha_\theta$, introduced in ~\cite{Giese:2020rtr},
\begin{align}
\label{eq:crit_PT_gen}
    \alpha_\vartheta &\equiv \frac{4D \vartheta}{3 w_+(T_+)}
    \equiv \frac{4\left( D e(T_+) - \frac{\delta e}{\delta p}(T_+,T_-) D p(T_+) \right)}{3 w_+(T_+)}\,,
\end{align}
where the $D$ and $\delta$ are defined as $D f = f_+(T_+)-f_-(T_+)$ and $\delta f = f_-(T_+)-f_-(T_-)$. For given values of $T_\pm$, they can be related to $v_\pm$ via the matching conditions in \eqref{eq:rel_velocities}, then, inverse hydrodynamics takes place for $\alpha_\vartheta <0$, while the standard one is realised for $\alpha_\vartheta > 0$. In this way, we discover that PTs proceeding against the vacuum energy can nonetheless display direct hydrodynamics.

Notice that for relatively weak PTs with $T_+ \simeq T_-$, $\delta e/\delta p \simeq 1/c_{s,-}^2$, with $c_{s,-}$ being the speed of sound in the broken phase, Eq.\,\eqref{eq:crit_PT_gen} reduces to $\alpha_\theta$ as defined in Ref.~\cite{Giese:2020rtr}. In the special case of a strictly constant speed of sound, one can refer to the template $\mu\nu$ model as introduced in Ref.\,\cite{Leitao:2014pda} to capture deviations from the relativistic fluid with $c_s^2 \neq 1/3$. In this case, our definition further reduces to $\alpha_\theta$ as derived within this template. 
Finally, when the speed of sound is $c_s^2 = 1/3$ as for a relativistic gas, this definition reduces to $\alpha_+$ as considered in Ref.~\cite{Barni:2024lkj}.

One can show that FOPTs with $\alpha_\vartheta = 0$ represent the limit of weak hydrodynamics, where $\Delta e= 0$ and $\Delta p=0$, with $\Delta f = f_+(T_+)-f_-(T_-)$. By continuity, this is supposed to separate inverse from direct FOPTs.
\begin{figure}
 \centering  \includegraphics[width=0.48\textwidth]{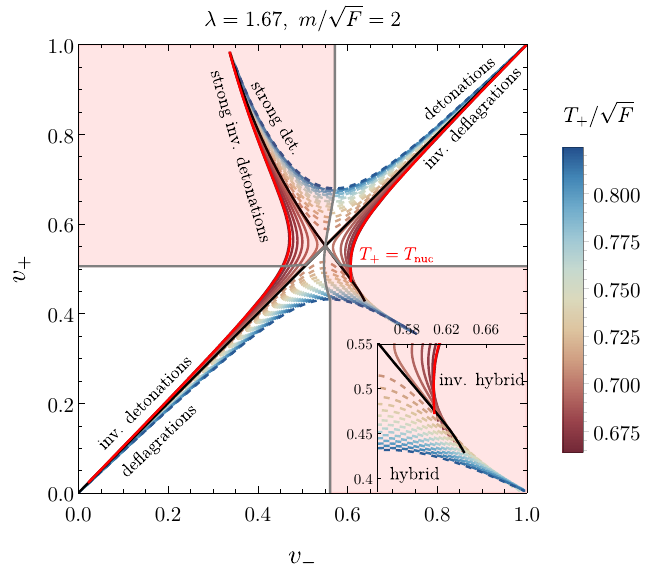}
 \caption{Possible solutions to the fluid matching conditions for \( (v_-, v_+) \) for the $R$--symmetry breaking FOPT under consideration, plotting the relevant branches for different values of $T_+$ between $T_c$ and the temperature where the barrier disappears.
 Dashed lines correspond to direct phase transitions, while solid lines indicate inverse transitions, as determined by the sign of \( \alpha_\vartheta \). The solid red line highlights the relevant branch at $T_{\text{nuc}}$. The red-shaded area marks the region of strong (inverse) detonations and strong (inverse) deflagrations. In the bottom right corner, a zoomed-in view of the hybrid solution region reveals an overlap between different branches (see Appendix\,\ref{app:hydro} for more details). 
}
 \label{fig:Full_branches}
 \end{figure}
 
\medskip
Let us now examine the possible hydrodynamics of the $R$--symmetry breaking FOPT. The junction conditions above can be solved numerically by referring to the pressure and energy densities as evaluated directly from the free energy within our particle physics model. The allowed values for the $(v_-,v_+)$ pairs are shown in Fig.\,\ref{fig:Full_branches} for a representative benchmark point. The matching conditions in Eq.\,\eqref{eq:rel_velocities} are solved for $v_\pm$ in terms of the temperatures ahead and behind the wall, $T_\pm$. For consistency, we restrict $T_+$ to lie between $T_c$ and the temperature when the barrier disappears, as this is the range for which the FOPT can actually take place. The various $v_\pm$ trajectories in Fig.\,\ref{fig:Full_branches} are then shown together with the corresponding temperature $T_+$ according to the colour code. Because of the consistency condition on $T_+$ and the properties of our system free energy, the branches do not populate the entire $v_\pm \in (0,1)$ parameter space. 
The regions corresponding to inverse and direct hydrodynamics, according to the sign of $\alpha_\vartheta$, are indicated by solid and dashed lines, respectively.
We find that these regions remain neatly separated across the entire $(v_-,v_+)$ plane, except for a small overlap in the regime of hybrid solutions (bottom-right corner). As a comparison, a similar discussion of the inverse branches in the case of the simplified (template) $\mu\nu$-model is provided in Appendix~\ref{app:munu}, where we find qualitative agreement with the full numerical study of the SUSY model presented here.

In the early universe, bubbles are efficiently formed when the nucleation rate catches up with the Hubble expansion. This condition, presented in more detail in App.\,\ref{app:dynamics} and Refs.\,\cite{Coleman:1977py,Linde:1980tt,Linde:1981zj,Ellis:2019oqb,PhysRevLett.44.963.2,Enqvist:1991xw}, connects the onset of the FOPT with a certain nucleation temperature, $T_{\rm{nuc}}$.
If we then further specify the temperature of the FOPT as $T_{\rm{nuc}}$\,\footnote{The relation $T_\text{nuc} = T_+$ only holds for detonations and anti--deflagrations. For the other expansion modes, we still use this as a sensible approximation to identify the relevant $v_\pm$ branch.}, we can select the bright red branch as the relevant one for this specific benchmark point. Notice that, as the matching conditions can not uniquely determine the bubble wall velocity, the actual value of $v_\pm$ cannot be pinned down by the hydrodynamics only, and the full red branch can in principle be realised. On the other hand, when taking the wall velocity as an additional input, the fluid profile can be fully determined. 
As we can see, the FOPT within this benchmark point occurs in the inverse hydrodynamic regime.

\begin{figure}
 \centering  
  \includegraphics[width=0.48\textwidth]{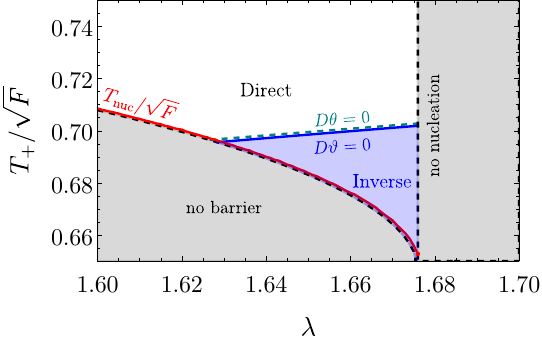}
 \caption{The nucleation temperature (red line) is obtained as a function of \( \lambda \) by numerically solving the condition \( S_3/T = 140 \), which corresponds to setting \( \sqrt{F} \sim \) TeV for concreteness (see App.\,\ref{app:dynamics}).
 The blue-shaded (white) region indicates the occurrence of the inverse (direct) FOPTs, whose boundary is shown according to the criteria \( D\vartheta = 0 \) and \( D\theta = 0 \). For this figure we fixed $m/\sqrt{F} = 2$.}
 \label{fig:Tnuc}
 \end{figure}

In Fig.\,\ref{fig:Tnuc} we perform a scan over the model parameter space, by fixing $m/\sqrt{F}=2$ and varying the coupling constant $\lambda$. The red line indicates the nucleation temperature, which always happens to be very close to the temperature where the barrier actually disappears. 

For $\lambda \lesssim 1.63$, bubble nucleation occurs in the region where the hydrodynamics will be the one based on the (direct) detonation and deflagration types of solutions, while for $1.63 \lesssim \lambda \lesssim 1.68$ the hydrodynamics will be inverse. We can also notice that the condition of vanishing $\alpha_\vartheta$ actually corresponds to the boundary between direct and inverse regions, which are determined independently by solving the fluid equations. As we can see, the approximate condition in terms of the pseudo-trace, $\alpha_\theta = 0$, reproduces this separation fairly well. This can be traced back to the fact that the speed of sound is not strongly temperature dependent in this model.

\medskip

\begin{figure*}
 \centering  
 \begin{minipage}{\textwidth}
 \centering \begin{minipage}{0.325\textwidth}
 \centering
     \includegraphics[width=\textwidth]{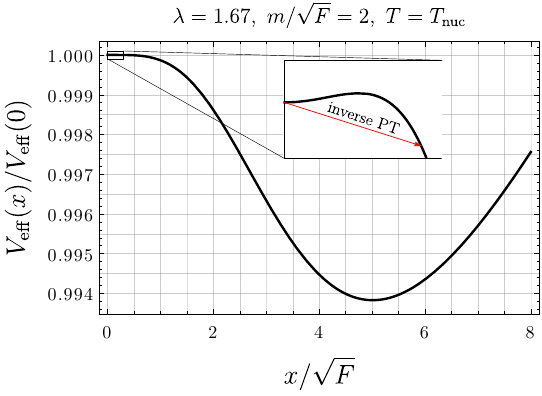}
 \end{minipage}\begin{minipage}{0.325\textwidth}
 \centering
     \includegraphics[width=0.7\textwidth]{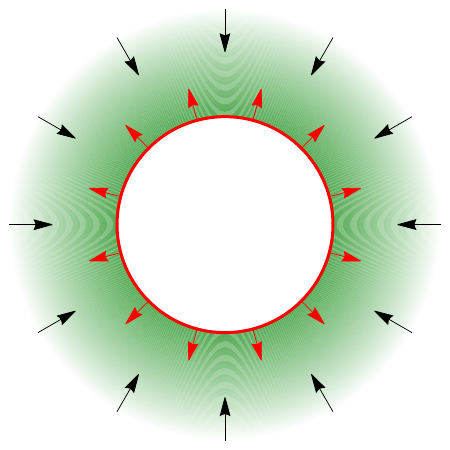}
 \end{minipage}\begin{minipage}{0.325\textwidth}
 \centering
 \includegraphics[width=\textwidth]{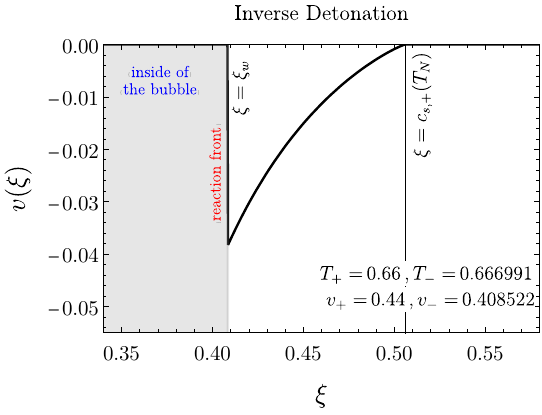}\end{minipage}\end{minipage}
 \caption{
Free energy of the system at finite temperature evaluated at one--loop at the nucleation temperature, with the arrow indicating the direction of the phase transition towards the minimum with a non--zero $x$ (left panel), together with a sketch of the expanding bubble and its velocity in red, and the fluid profile in green (central panel) for a characteristic inverse detonation.
The actual fluid profile in the plasma frame is shown in the right panel. Due to its inverse nature, the fluid velocity is always negative. See main text and Appendix~\ref{app:hydro} for details.
 }
 \label{fig:schematic}
 \end{figure*}

{\it Inverse fluid solutions for $R$--symmetry breaking --} The hydrodynamics of inverse PTs was presented for the first time in Ref.~\cite{Barni:2024lkj, Bea:2024bxu} (see also\,\cite{Bea:2024bls}):
there exist five different possible expansion modes with negative bulk velocities: i) inverse detonations (weak and Chapman-Jouguet (CJ)), ii) inverse deflagrations (weak and CJ), and iii) inverse hybrids.

This classification of hydrodynamic solutions was obtained within the (simplified) bag EoS. We have checked that this picture remains qualitatively the same also when considering the full form of the free energy (or effective potential) as evaluated explicitly for the SUSY model under consideration. In practice, we find only some quantitative differences related to the actual value of the speed of sound, which generally differs from $c_s^2 = 1/3$, and to the (mild) temperature dependence of $c_s^2$, which requires solving the coupled system of fluid equations for the pressure and the energy density as discussed in Appendix \,\eqref{app:hydro}. An example of the explicit profiles obtained by solving numerically the fluid equations for the benchmark point with $\lambda=1.67$ and $m/\sqrt{F} = 2$ are shown in detail in Fig.\,\ref{fig:schematic} for an inverse detonation, together with the free energy of the system at the nucleation temperature showing the direction of the phase transition and a sketch of the bubble with the corresponding fluid profile.

Let us also mention that there is in principle the possibility that the bubble wall never reaches any of the steady states presented above, and keeps accelerating until bubbles collide, namely it runs away.
Employing the line of reasoning presented in Ref.~\cite{Barni:2024lkj}, we find that the bubble never runs away in the model under consideration, and always reaches one of the steady states (see App.,\ref{app:velo} and Refs.\,\cite{Ai:2023see,Ai:2024uyw,Ai:2024shx} for a derivation). Our hydrodynamic analysis however cannot determine which one of them, as mentioned above.

{\it Coupling to the SM thermal bath --}
In the early universe, the SUSY breaking sector considered here is generally accompanied by additional spectator fields~\footnote{A spectator field does not change its physical properties, \emph{e.g.} its mass, during the phase transition.} that are in thermal equilibrium with the SUSY breaking sector and constitute a radiation bath. 
To assess the impact of these additional degrees of freedom, we redefine the energy and pressure as
\begin{equation}
    p(T) \to p(T) + \tilde c^2 \tilde{a} T^4, \quad e(T) \to e(T) + 3 \tilde c^2\tilde{a} T^4 \, ,
\end{equation}
where $\tilde c^2 =1/3$, and \(\tilde{a}\) controls the number of the relativistic spectator degrees of freedom (dofs), which is expected to be $\tilde{a}  \sim 70$ considering a supersymmetric extension of the Standard Model. 

The presence of these fields will mostly influence the strength of the FOPT. In the limit $\tilde a \gg 1$, one has $\delta p/\delta e \simeq 1/3$ as expected for a gas of relativistic particles, and the generalised pseudo-trace in this limit becomes
\bea 
\label{eq:spectator}
\alpha_\vartheta \simeq \frac{4 (D e - 3D p)}{3 w_+(T_+)} \frac{1}{1+x}\,, 
\quad x = \frac{4 \tilde{c}^2 \tilde{a} T_+^4}{w_+(T_+)}\,.
\eea
Thus, to a good approximation, the strength of the phase transition exhibits an inverse scaling with \( \tilde{a} \), aligning with physical intuition.
From explicit calculations, we find that the pseudo-trace and generalised pseudo-trace are always very close to each other in the parameter space of interest, and that the asymptotic behaviour in Eq.\,\eqref{eq:spectator} is well established for $\tilde a \gtrsim 50$ leading to typical values of $\alpha_\vartheta \lesssim 10^{-2}$, while in the absence of spectator fields one would have $\alpha_\vartheta \lesssim 10^{-1}$.

In this regard, let us notice that there is in fact a fundamental difference between the strength of a standard (direct) FOPT and the case of an inverse FOPT. By referring to the definition of $\alpha_\vartheta$ in Eq.\,\eqref{eq:crit_PT_gen}, we can see that the part containing $Dp(T_+)$ will always contribute with a positive sign. This follows from the fact that the broken phase will necessarily have a larger pressure than the symmetric phase for the FOPT to take place and that $\delta e/\delta p \simeq 1/c_s^2$ is a positive quantity. Therefore, considering the case of negative $\alpha_\vartheta$, we can derive the following inequality:
\begin{equation}
\label{eq:ineq}
    \frac{3}{4}|\alpha_\vartheta| <  \frac{\omega_-(T_+) - \omega_+(T_+)}{\omega_+(T_+)} = \frac{\Delta a_{\text{eff}}(T_+)}{a_{{\text{eff}},+}(T_+)}\,,
\end{equation}
where $a_{{\text{eff}},+}(T_+)$ indicates the effective number of relativistic dofs in the symmetric phase at the temperature $T_+$, according to the parametrization $3\omega(T)/4T^4 \equiv a_{\text{eff}}(T)$, and $\Delta a_{\text{eff}}(T_+)$ is the change in dofs in the broken phase at the same temperature. This relation shows that an inverse FOPT can be strong only when it involves a significant change in dofs between the two phases. This is a structural property of the vacua of the theory under consideration, and it should be contrasted with the case of standard FOPTs whose strength is mostly controlled by the amount of supercooling that can be achieved in the expanding universe. In particular, Eq.\,\eqref{eq:ineq} indicates that an inverse FOPT is not necessarily stronger when it becomes more supercooled. 

{\it Conclusion and outlook --}
We presented a simple SUSY breaking model displaying a window of \emph{inverse} FOPTs during the spontaneous breaking of the $R$--symmetry. This represents the first explicit example of a BSM model leading to an inverse FOPT in a cooling cosmology, as well as a proof of principle for the relevance of this dynamics in the early universe. 

We find that the sign of the generalised pseudo-trace, $\alpha_\vartheta$ in Eq.\eqref{eq:crit_PT_gen}, determines the \emph{inverseness} of the transition. As a comparison, we also show that the sign of the pseudo-trace introduced in Ref.\,\cite{Giese:2020rtr} offers a fair estimate for the type of the FOPT as well. 

Our study motivates a broader investigation of inverse FOPTs in explicit BSM models. This includes establishing a deeper connection between the inverseness of a FOPT and its fundamental properties and symmetries, exemplified here within a model of spontaneous SUSY breaking, as well as identifying possible non--SUSY realisations of this dynamics.

Finally, FOPTs are powerful sources of gravitational waves that can be detected at current and forthcoming GW observatories. This work provides motivation to characterize the GW spectrum related to inverse FOPTs, and to determine to which extent this can be distinguished from the one arising during direct FOPTs.
 
{\it Acknowledgments--}
We sincerely acknowledge Thomas Konstandin, Xander Nagels, Alberto Mariotti, David Mateos, Diego Redigolo, and Mikel Sanchez-Garitaonandia for clarifying discussions and for reading the manuscript. 
GB is supported by the grant CNS2023-145069 funded by MICIU/AEI/10.13039/501100011033 and by the European Union NextGenerationEU/PRTR. GB also acknowledges the support of the Spanish Agencia Estatal de Investigacion through the grant “IFT Centro de Excelencia Severo Ochoa CEX2020-001007-S”.
SB is supported by the Deutsche Forschungsgemeinschaft under Germany’s Excellence Strategy - EXC 2121 Quantum universe - 390833306. MV is supported by the ``Excellence of Science - EOS'' - be.h project n.30820817, and by the Strategic Research Program High-Energy Physics of the Vrije Universiteit Brussel.

\bibliography{biblio}

@article{Canko:2017ebb,
    author = "Canko, D. and Gialamas, I. and Jelic-Cizmek, G. and Riotto, A. and Tetradis, N.",
    title = "{On the Catalysis of the Electroweak Vacuum Decay by Black Holes at High Temperature}",
    eprint = "1706.01364",
    archivePrefix = "arXiv",
    primaryClass = "hep-th",
    doi = "10.1140/epjc/s10052-018-5808-y",
    journal = "Eur. Phys. J. C",
    volume = "78",
    number = "4",
    pages = "328",
    year = "2018"
}

@article{Dent:2024bhi,
    author = "Dent, James B. and Dutta, Bhaskar and Rai, Mudit",
    title = "{Imprints of Early Universe Cosmology on Gravitational Waves}",
    eprint = "2411.09757",
    archivePrefix = "arXiv",
    primaryClass = "hep-ph",
    month = "11",
    year = "2024"
}

@article{Buen-Abad:2023hex,
    author = "Buen-Abad, Manuel A. and Chang, Jae Hyeok and Hook, Anson",
    title = "{Gravitational wave signatures from reheating}",
    eprint = "2305.09712",
    archivePrefix = "arXiv",
    primaryClass = "hep-ph",
    doi = "10.1103/PhysRevD.108.036006",
    journal = "Phys. Rev. D",
    volume = "108",
    number = "3",
    pages = "036006",
    year = "2023"
}

@article{Ai:2023see,
    author = "Ai, Wen-Yuan and Laurent, Benoit and van de Vis, Jorinde",
    title = "{Model-independent bubble wall velocities in local thermal equilibrium}",
    eprint = "2303.10171",
    archivePrefix = "arXiv",
    primaryClass = "astro-ph.CO",
    reportNumber = "KCL-PH-TH/2023-19",
    doi = "10.1088/1475-7516/2023/07/002",
    journal = "JCAP",
    volume = "07",
    pages = "002",
    year = "2023"
}

@article{Ai:2024shx,
    author = "Ai, Wen-Yuan and Nagels, Xander and Vanvlasselaer, Miguel",
    title = "{Criterion for ultra-fast bubble walls: the impact of hydrodynamic obstruction}",
    eprint = "2401.05911",
    archivePrefix = "arXiv",
    primaryClass = "hep-ph",
    reportNumber = "KCL-PH-TH/2024-01",
    doi = "10.1088/1475-7516/2024/03/037",
    journal = "JCAP",
    volume = "03",
    pages = "037",
    year = "2024"
}

@article{Ai:2021kak,
    author = "Ai, Wen-Yuan and Garbrecht, Bjorn and Tamarit, Carlos",
    title = "{Bubble wall velocities in local equilibrium}",
    eprint = "2109.13710",
    archivePrefix = "arXiv",
    primaryClass = "hep-ph",
    reportNumber = "CP3-21-53, TUM-HEP-1365-21",
    doi = "10.1088/1475-7516/2022/03/015",
    journal = "JCAP",
    volume = "03",
    number = "03",
    pages = "015",
    year = "2022"
}

@article{Laine:1993ey,
    author = "Laine, M.",
    title = "{Bubble growth as a detonation}",
    eprint = "hep-ph/9309242",
    archivePrefix = "arXiv",
    reportNumber = "HU-TFT-93-44",
    doi = "10.1103/PhysRevD.49.3847",
    journal = "Phys. Rev. D",
    volume = "49",
    pages = "3847--3853",
    year = "1994"
}

@article{Kurki-Suonio:1995rrv,
    author = "Kurki-Suonio, H. and Laine, M.",
    title = "{Supersonic deflagrations in cosmological phase transitions}",
    eprint = "hep-ph/9501216",
    archivePrefix = "arXiv",
    reportNumber = "HU-TFT-95-3",
    doi = "10.1103/PhysRevD.51.5431",
    journal = "Phys. Rev. D",
    volume = "51",
    pages = "5431--5437",
    year = "1995"
}

@article{Sanchez-Garitaonandia:2023zqz,
    author = "Sanchez-Garitaonandia, Mikel and van de Vis, Jorinde",
    title = "{Prediction of the bubble wall velocity for a large jump in degrees of freedom}",
    eprint = "2312.09964",
    archivePrefix = "arXiv",
    primaryClass = "hep-ph",
    month = "12",
    year = "2023"
}

@article{Giese:2020rtr,
    author = "Giese, Felix and Konstandin, Thomas and van de Vis, Jorinde",
    title = "{Model-independent energy budget of cosmological first-order phase transitions\textemdash{}A sound argument to go beyond the bag model}",
    eprint = "2004.06995",
    archivePrefix = "arXiv",
    primaryClass = "astro-ph.CO",
    reportNumber = "DESY-20-064",
    doi = "10.1088/1475-7516/2020/07/057",
    journal = "JCAP",
    volume = "07",
    number = "07",
    pages = "057",
    year = "2020"
}

@article{Casalderrey-Solana:2022rrn,
    author = "Casalderrey-Solana, Jorge and Mateos, David and Sanchez-Garitaonandia, Mikel",
    title = "{Mega-Hertz Gravitational Waves from Neutron Star Mergers}",
    eprint = "2210.03171",
    archivePrefix = "arXiv",
    primaryClass = "hep-th",
    month = "10",
    year = "2022"
}

@article{Hong:2020est,
    author = "Hong, Jeong-Pyong and Jung, Sunghoon and Xie, Ke-Pan",
    title = "{Fermi-ball dark matter from a first-order phase transition}",
    eprint = "2008.04430",
    archivePrefix = "arXiv",
    primaryClass = "hep-ph",
    doi = "10.1103/PhysRevD.102.075028",
    journal = "Phys. Rev. D",
    volume = "102",
    number = "7",
    pages = "075028",
    year = "2020"
}

@article{Giese:2020znk,
    author = "Giese, Felix and Konstandin, Thomas and Schmitz, Kai and van de Vis, Jorinde",
    title = "{Model-independent energy budget for LISA}",
    eprint = "2010.09744",
    archivePrefix = "arXiv",
    primaryClass = "astro-ph.CO",
    reportNumber = "DESY-20-173, DESY 20-173, CERN-TH-2020-170",
    doi = "10.1088/1475-7516/2021/01/072",
    journal = "JCAP",
    volume = "01",
    pages = "072",
    year = "2021"
}

@article{Halverson:2020xpg,
    author = "Halverson, James and Long, Cody and Maiti, Anindita and Nelson, Brent and Salinas, Gustavo",
    title = "{Gravitational waves from dark Yang-Mills sectors}",
    eprint = "2012.04071",
    archivePrefix = "arXiv",
    primaryClass = "hep-ph",
    doi = "10.1007/JHEP05(2021)154",
    journal = "JHEP",
    volume = "05",
    pages = "154",
    year = "2021"
}

@article{Ghosh:2020ipy,
    author = "Ghosh, Tathagata and Guo, Huai-Ke and Han, Tao and Liu, Hongkai",
    title = "{Electroweak phase transition with an SU(2) dark sector}",
    eprint = "2012.09758",
    archivePrefix = "arXiv",
    primaryClass = "hep-ph",
    doi = "10.1007/JHEP07(2021)045",
    journal = "JHEP",
    volume = "07",
    pages = "045",
    year = "2021"
}

@article{Aoki:2021oez,
    author = "Aoki, Mayumi and Komatsu, Takatoshi and Shibuya, Hiroto",
    title = "{Possibility of a multi-step electroweak phase transition in the two-Higgs doublet models}",
    eprint = "2106.03439",
    archivePrefix = "arXiv",
    primaryClass = "hep-ph",
    reportNumber = "KANAZAWA-21-08",
    doi = "10.1093/ptep/ptac068",
    journal = "PTEP",
    volume = "2022",
    number = "6",
    pages = "063B05",
    year = "2022"
}

@article{Banerjee:2024qiu,
    author = "Banerjee, Upalaparna and Chakraborty, Sabyasachi and Prakash, Suraj and Rahaman, Shakeel Ur",
    title = "{The feasibility of ultra-relativistic bubbles in SMEFT}",
    eprint = "2402.02914",
    archivePrefix = "arXiv",
    primaryClass = "hep-ph",
    month = "2",
    year = "2024"
}

@article{Lu:2022paj,
    author = "Lu, Philip and Kawana, Kiyoharu and Xie, Ke-Pan",
    title = "{Old phase remnants in first-order phase transitions}",
    eprint = "2202.03439",
    archivePrefix = "arXiv",
    primaryClass = "astro-ph.CO",
    doi = "10.1103/PhysRevD.105.123503",
    journal = "Phys. Rev. D",
    volume = "105",
    number = "12",
    pages = "123503",
    year = "2022"
}

@article{Kierkla:2022odc,
    author = "Kierkla, Maciej and Karam, Alexandros and Swiezewska, Bogumila",
    title = "{Conformal model for gravitational waves and dark matter: a status update}",
    eprint = "2210.07075",
    archivePrefix = "arXiv",
    primaryClass = "astro-ph.CO",
    doi = "10.1007/JHEP03(2023)007",
    journal = "JHEP",
    volume = "03",
    pages = "007",
    year = "2023"
}

@article{Reichert:2022naa,
    author = "Reichert, Manuel and Wang, Zhi-Wei",
    title = "{Gravitational Waves from dark composite dynamics}",
    eprint = "2211.08877",
    archivePrefix = "arXiv",
    primaryClass = "hep-ph",
    doi = "10.1051/epjconf/202227408003",
    journal = "EPJ Web Conf.",
    volume = "274",
    pages = "08003",
    year = "2022"
}

@article{Badziak:2022ltm,
    author = "Badziak, Marcin and Nalecz, Ignacy",
    title = "{First-order phase transitions in Twin Higgs models}",
    eprint = "2212.09776",
    archivePrefix = "arXiv",
    primaryClass = "hep-ph",
    doi = "10.1007/JHEP02(2023)185",
    journal = "JHEP",
    volume = "02",
    pages = "185",
    year = "2023"
}

@article{Frandsen:2023vhu,
    author = "Frandsen, Mads T. and Heikinheimo, Matti and Rosenlyst, Martin and Thing, Mattias E. and Tuominen, Kimmo",
    title = "{Gravitational waves from SU(N)/SP(N) composite Higgs models}",
    eprint = "2302.09104",
    archivePrefix = "arXiv",
    primaryClass = "hep-ph",
    doi = "10.1007/JHEP09(2023)022",
    journal = "JHEP",
    volume = "09",
    pages = "022",
    year = "2023"
}

@article{Addazi:2023ftv,
    author = "Addazi, Andrea and Marcian\`o, Antonino and Morais, Ant\'onio P. and Pasechnik, Roman and Viana, Jo\~ao and Yang, Hao",
    title = "{Gravitational echoes of lepton number symmetry breaking with light and ultralight Majorons}",
    eprint = "2304.02399",
    archivePrefix = "arXiv",
    primaryClass = "hep-ph",
    reportNumber = "CERN-TH-2023-054",
    doi = "10.1088/1475-7516/2023/09/026",
    journal = "JCAP",
    volume = "09",
    pages = "026",
    year = "2023",
    note = "[Erratum: JCAP 03, E01 (2024)]"
}

@article{Fujikura:2023fbi,
    author = "Fujikura, Kohei and Nakai, Yuichiro and Sato, Ryosuke and Wang, Yaoduo",
    title = "{Cosmological phase transitions in composite Higgs models}",
    eprint = "2306.01305",
    archivePrefix = "arXiv",
    primaryClass = "hep-ph",
    reportNumber = "OU-HET-1190, UT-Komaba/23-5",
    doi = "10.1007/JHEP09(2023)053",
    journal = "JHEP",
    volume = "09",
    pages = "053",
    year = "2023"
}

@article{Jinno:2016knw,
    author = "Jinno, Ryusuke and Takimoto, Masahiro",
    title = "{Probing a classically conformal B-L model with gravitational waves}",
    eprint = "1604.05035",
    archivePrefix = "arXiv",
    primaryClass = "hep-ph",
    reportNumber = "KEK-TH-1896",
    doi = "10.1103/PhysRevD.95.015020",
    journal = "Phys. Rev. D",
    volume = "95",
    number = "1",
    pages = "015020",
    year = "2017"
}

@article{Morgante:2022zvc,
    author = "Morgante, Enrico and Ramberg, Nicklas and Schwaller, Pedro",
    title = "{Gravitational waves from dark SU(3) Yang-Mills theory}",
    eprint = "2210.11821",
    archivePrefix = "arXiv",
    primaryClass = "hep-ph",
    doi = "10.1103/PhysRevD.107.036010",
    journal = "Phys. Rev. D",
    volume = "107",
    number = "3",
    pages = "036010",
    year = "2023"
}

@article{Pasechnik:2023hwv,
    author = "Pasechnik, Roman and Reichert, Manuel and Sannino, Francesco and Wang, Zhi-Wei",
    title = "{Gravitational waves from composite dark sectors}",
    eprint = "2309.16755",
    archivePrefix = "arXiv",
    primaryClass = "hep-ph",
    doi = "10.1007/JHEP02(2024)159",
    journal = "JHEP",
    volume = "02",
    pages = "159",
    year = "2024"
}

@article{Giudice:2024tcp,
    author = "Giudice, Gian F. and Lee, Hyun Min and Pomarol, Alex and Shakya, Bibhushan",
    title = "{Nonthermal Heavy Dark Matter from a First-Order Phase Transition}",
    eprint = "2403.03252",
    archivePrefix = "arXiv",
    primaryClass = "hep-ph",
    reportNumber = "CERN-TH-2024-031, DESY-24-031",
    month = "3",
    year = "2024"
}

@article{Hogan_GW_1986,
    author = "Hogan, C. J.",
    title = "{Gravitational radiation from cosmological phase transitions}",
    journal = "Mon. Not. Roy. Astron. Soc.",
    volume = "218",
    pages = "629--636",
    year = "1986"
}

@article{Laine:1998jb,
    author = "Laine, M. and Rummukainen, K.",
    editor = "DeGrand, Thomas A. and DeTar, Carleton E. and Sugar, R. and Toussaint, D.",
    title = "{What's new with the electroweak phase transition?}",
    eprint = "hep-lat/9809045",
    archivePrefix = "arXiv",
    doi = "10.1016/S0920-5632(99)85017-8",
    journal = "Nucl. Phys. B Proc. Suppl.",
    volume = "73",
    pages = "180--185",
    year = "1999"
}

@article{Baldes:2021vyz,
    author = "Baldes, Iason and Blasi, Simone and Mariotti, Alberto and Sevrin, Alexander and Turbang, Kevin",
    title = "{Baryogenesis via relativistic bubble expansion}",
    eprint = "2106.15602",
    archivePrefix = "arXiv",
    primaryClass = "hep-ph",
    reportNumber = "ULB-TH/21-09",
    doi = "10.1103/PhysRevD.104.115029",
    journal = "Phys. Rev. D",
    volume = "104",
    number = "11",
    pages = "115029",
    year = "2021"
}

@article{Delaunay:2007wb,
    author = "Delaunay, Cedric and Grojean, Christophe and Wells, James D.",
    title = "{Dynamics of Non-renormalizable Electroweak Symmetry Breaking}",
    eprint = "0711.2511",
    archivePrefix = "arXiv",
    primaryClass = "hep-ph",
    reportNumber = "CERN-PH-TH-2007-219, MCTP-07-31, SACLAY-T07-141",
    doi = "10.1088/1126-6708/2008/04/029",
    journal = "JHEP",
    volume = "04",
    pages = "029",
    year = "2008"
}

@article{Nelson:1991ab,
    author = "Nelson, A. E. and Kaplan, D. B. and Cohen, Andrew G.",
    title = "{Why there is something rather than nothing: Matter from weak interactions}",
    reportNumber = "UCSD-PTH-91-20, BUHEP-91-15",
    doi = "10.1016/0550-3213(92)90440-M",
    journal = "Nucl. Phys. B",
    volume = "373",
    pages = "453--478",
    year = "1992"
}

@article{Carena:1996wj,
    author = "Carena, Marcela and Quiros, M. and Wagner, C. E. M.",
    title = "{Opening the window for electroweak baryogenesis}",
    eprint = "hep-ph/9603420",
    archivePrefix = "arXiv",
    reportNumber = "CERN-TH-96-30, IEM-FT-126-96",
    doi = "10.1016/0370-2693(96)00475-3",
    journal = "Phys. Lett. B",
    volume = "380",
    pages = "81--91",
    year = "1996"
}

@article{Baldes:2020kam,
    author = "Baldes, Iason and Gouttenoire, Yann and Sala, Filippo",
    title = "{String Fragmentation in Supercooled Confinement and Implications for Dark Matter}",
    eprint = "2007.08440",
    archivePrefix = "arXiv",
    primaryClass = "hep-ph",
    reportNumber = "DESY-20-122",
    doi = "10.1007/JHEP04(2021)278",
    journal = "JHEP",
    volume = "04",
    pages = "278",
    year = "2021"
}

@article{Azatov:2021ifm,
    author = "Azatov, Aleksandr and Vanvlasselaer, Miguel and Yin, Wen",
    title = "{Dark Matter production from relativistic bubble walls}",
    eprint = "2101.05721",
    archivePrefix = "arXiv",
    primaryClass = "hep-ph",
    reportNumber = "SISSA 03/2021/FISI",
    doi = "10.1007/JHEP03(2021)288",
    journal = "JHEP",
    volume = "03",
    pages = "288",
    year = "2021"
}

@article{Azatov:2019png,
    author = "Azatov, Aleksandr and Barducci, Daniele and Sgarlata, Francesco",
    title = "{Gravitational traces of broken gauge symmetries}",
    eprint = "1910.01124",
    archivePrefix = "arXiv",
    primaryClass = "hep-ph",
    reportNumber = "SISSA 28/2019/FISI",
    doi = "10.1088/1475-7516/2020/07/027",
    journal = "JCAP",
    volume = "07",
    pages = "027",
    year = "2020"
}

@article{Shaposhnikov:1986jp,
    author = "Shaposhnikov, M.E.",
    title = "{Possible Appearance of the Baryon Asymmetry of the Universe in an Electroweak Theory}",
    journal = "JETP Lett.",
    volume = "44",
    pages = "465--468",
    year = "1986"
}

@article{Bruggisser:2018mus,
    author = "Bruggisser, Sebastian and Von Harling, Benedict and Matsedonskyi, Oleksii and Servant, Géraldine",
    title = "{Baryon Asymmetry from a Composite Higgs Boson}",
    eprint = "1803.08546",
    archivePrefix = "arXiv",
    primaryClass = "hep-ph",
    reportNumber = "DESY-18-029",
    doi = "10.1103/PhysRevLett.121.131801",
    journal = "Phys. Rev. Lett.",
    volume = "121",
    number = "13",
    pages = "131801",
    year = "2018"
}

@article{Bruggisser:2018mrt,
    author = "Bruggisser, Sebastian and Von Harling, Benedict and Matsedonskyi, Oleksii and Servant, Géraldine",
    title = "{Electroweak Phase Transition and Baryogenesis in Composite Higgs Models}",
    eprint = "1804.07314",
    archivePrefix = "arXiv",
    primaryClass = "hep-ph",
    reportNumber = "DESY-17-229",
    doi = "10.1007/JHEP12(2018)099",
    journal = "JHEP",
    volume = "12",
    pages = "099",
    year = "2018"
}

@article{DelleRose:2019pgi,
    author = "Delle Rose, Luigi and Panico, Giuliano and Redi, Michele and Tesi, Andrea",
    title = "{Gravitational Waves from Supercool Axions}",
    eprint = "1912.06139",
    archivePrefix = "arXiv",
    primaryClass = "hep-ph",
    doi = "10.1007/JHEP04(2020)025",
    journal = "JHEP",
    volume = "04",
    pages = "025",
    year = "2020"
}

@article{Espinosa:2010hh,
      author         = "Espinosa, Jose R. and Konstandin, Thomas and No, Jose M.
                        and Servant, Geraldine",
      title          = "{Energy Budget of Cosmological First-order Phase
                        Transitions}",
      journal        = "JCAP",
      volume         = "1006",
      year           = "2010",
      pages          = "028",
      doi            = "10.1088/1475-7516/2010/06/028",
      eprint         = "1004.4187",
      archivePrefix  = "arXiv",
      primaryClass   = "hep-ph",
      reportNumber   = "CERN-PH-TH-2010-027",
      SLACcitation   = "%%CITATION = ARXIV:1004.4187;%%"
}

@article{Curtin:2016urg,
      author         = "Curtin, David and Meade, Patrick and Ramani,
                        Harikrishnan",
      title          = "{Thermal Resummation and Phase Transitions}",
      journal        = "Eur. Phys. J.",
      volume         = "C78",
      year           = "2018",
      number         = "9",
      pages          = "787",
      doi            = "10.1140/epjc/s10052-018-6268-0",
      eprint         = "1612.00466",
      archivePrefix  = "arXiv",
      primaryClass   = "hep-ph",
      reportNumber   = "YITP-2016-48",
      SLACcitation   = "%%CITATION = ARXIV:1612.00466;%%"
}

@article{Caprini:2015zlo,
      author         = "Caprini, Chiara and others",
      title          = "{Science with the space-based interferometer eLISA. II:
                        Gravitational waves from cosmological phase transitions}",
      journal        = "JCAP",
      volume         = "1604",
      year           = "2016",
      number         = "04",
      pages          = "001",
      doi            = "10.1088/1475-7516/2016/04/001",
      eprint         = "1512.06239",
      archivePrefix  = "arXiv",
      primaryClass   = "astro-ph.CO",
      reportNumber   = "DESY-15-246",
      SLACcitation   = "%%CITATION = ARXIV:1512.06239;%%"
}

@article{Ellis:2019oqb,
      author         = "Ellis, John and Lewicki, Marek and No, Jos� Miguel and
                        Vaskonen, Ville",
      title          = "{Gravitational wave energy budget in strongly supercooled
                        phase transitions}",
      journal        = "JCAP",
      volume         = "1906",
      year           = "2019",
      number         = "06",
      pages          = "024",
      doi            = "10.1088/1475-7516/2019/06/024",
      eprint         = "1903.09642",
      archivePrefix  = "arXiv",
      primaryClass   = "hep-ph",
      reportNumber   = "KCL-PH-TH/2019-32, CERN-TH-2019-032, IFT-UAM/CSIC-19-32",
      SLACcitation   = "%%CITATION = ARXIV:1903.09642;%%"
}

@article{vonHarling:2017yew,
      author         = "von Harling, Benedict and Servant, Geraldine",
      title          = "{QCD-induced Electroweak Phase Transition}",
      journal        = "JHEP",
      volume         = "01",
      year           = "2018",
      pages          = "159",
      doi            = "10.1007/JHEP01(2018)159",
      eprint         = "1711.11554",
      archivePrefix  = "arXiv",
      primaryClass   = "hep-ph",
      reportNumber   = "DESY-17-056",
      SLACcitation   = "%%CITATION = ARXIV:1711.11554;%%"
}

@article{Coleman:1977py,
      author         = "Coleman, Sidney R.",
      title          = "{The Fate of the False Vacuum. 1. Semiclassical Theory}",
      journal        = "Phys. Rev.",
      volume         = "D15",
      year           = "1977",
      pages          = "2929-2936",
      doi            = "10.1103/PhysRevD.15.2929, 10.1103/PhysRevD.16.1248",
      note           = "[Erratum: Phys. Rev.D16,1248(1977)]",
      reportNumber   = "HUTP-77-A004",
      SLACcitation   = "%%CITATION = PHRVA,D15,2929;%%"
}

@article{PhysRevLett.44.963.2,
  title = {Phase Transitions and Magnetic Monopole Production in the Very Early Universe},
  author = {Guth, Alan H. and Tye, S. -H. H.},
  journal = {Phys. Rev. Lett.},
  volume = {44},
  issue = {14},
  pages = {963--963},
  numpages = {0},
  year = {1980},
  month = {Apr},
  publisher = {American Physical Society},
  doi = {10.1103/PhysRevLett.44.963.2},
  url = {https://link.aps.org/doi/10.1103/PhysRevLett.44.963.2}
}

@article{Baldes:2021aph,
    author = "Baldes, Iason and Gouttenoire, Yann and Sala, Filippo and Servant, G\'eraldine",
    title = "{Supercool composite Dark Matter beyond 100 TeV}",
    eprint = "2110.13926",
    archivePrefix = "arXiv",
    primaryClass = "hep-ph",
    reportNumber = "ULB-TH/21-17; DESY 21-172, ULB-TH/21-17, DESY 21-172",
    doi = "10.1007/JHEP07(2022)084",
    journal = "JHEP",
    volume = "07",
    pages = "084",
    year = "2022"
}

@article{Kodama:1982sf,
    author = "Kodama, Hideo and Sasaki, Misao and Sato, Katsuhiko",
    title = "{Abundance of Primordial Holes Produced by Cosmological First Order Phase Transition}",
    reportNumber = "KUNS 642",
    doi = "10.1143/PTP.68.1979",
    journal = "Prog. Theor. Phys.",
    volume = "68",
    pages = "1979",
    year = "1982"
}

@article{Baldes:2022oev,
    author = "Baldes, Iason and Gouttenoire, Yann and Sala, Filippo",
    title = "{Hot and heavy dark matter from a weak scale phase transition}",
    eprint = "2207.05096",
    archivePrefix = "arXiv",
    primaryClass = "hep-ph",
    reportNumber = "ULB-TH/22-12",
    doi = "10.21468/SciPostPhys.14.3.033",
    journal = "SciPost Phys.",
    volume = "14",
    pages = "033",
    year = "2023"
}

@article{Chun:2023ezg,
    author = "Chun, Eung Jin and Dutka, Tomasz P. and Jung, Tae Hyun and Nagels, Xander and Vanvlasselaer, Miguel",
    title = "{Bubble-assisted Leptogenesis}",
    eprint = "2305.10759",
    archivePrefix = "arXiv",
    primaryClass = "hep-ph",
    reportNumber = "CTPU-PTC-23-17",
    month = "5",
    year = "2023"
}

@article{Asadi:2021pwo,
    author = "Asadi, Pouya and Kramer, Eric David and Kuflik, Eric and Ridgway, Gregory W. and Slatyer, Tracy R. and Smirnov, Juri",
    title = "{Thermal squeezeout of dark matter}",
    eprint = "2103.09827",
    archivePrefix = "arXiv",
    primaryClass = "hep-ph",
    doi = "10.1103/PhysRevD.104.095013",
    journal = "Phys. Rev. D",
    volume = "104",
    number = "9",
    pages = "095013",
    year = "2021"
}

@article{Blasi:2022woz,
    author = "Blasi, Simone and Mariotti, Alberto",
    title = "{Domain Walls Seeding the Electroweak Phase Transition}",
    eprint = "2203.16450",
    archivePrefix = "arXiv",
    primaryClass = "hep-ph",
    doi = "10.1103/PhysRevLett.129.261303",
    journal = "Phys. Rev. Lett.",
    volume = "129",
    number = "26",
    pages = "261303",
    year = "2022"
}

@article{Agrawal:2023cgp,
    author = "Agrawal, Prateek and Blasi, Simone and Mariotti, Alberto and Nee, Michael",
    title = "{Electroweak phase transition with a double well done doubly well}",
    eprint = "2312.06749",
    archivePrefix = "arXiv",
    primaryClass = "hep-ph",
    reportNumber = "DESY-23-208",
    doi = "10.1007/JHEP06(2024)089",
    journal = "JHEP",
    volume = "06",
    pages = "089",
    year = "2024"
}

@article{Blasi:2024mtc,
    author = "Blasi, Simone and Mariotti, Alberto",
    title = "{QCD Axion Strings or Seeds?}",
    eprint = "2405.08060",
    archivePrefix = "arXiv",
    primaryClass = "hep-ph",
    doi = "10.21468/SciPostPhys.18.1.016",
    journal = "SciPost Phys.",
    volume = "18",
    pages = "016",
    year = "2025"
}

@article{Steinhardt:1981ec,
    author = "Steinhardt, Paul Joseph",
    title = "{Monopole and Vortex Dissociation and Decay of the False Vacuum}",
    reportNumber = "HUTP-80/A088",
    doi = "10.1016/0550-3213(81)90449-1",
    journal = "Nucl. Phys. B",
    volume = "190",
    pages = "583--616",
    year = "1981"
}

@article{Hosotani:1982ii,
    author = "Hosotani, Yutaka",
    title = "{Impurities in the Early Universe}",
    reportNumber = "UPR-0205T",
    doi = "10.1103/PhysRevD.27.789",
    journal = "Phys. Rev. D",
    volume = "27",
    pages = "789",
    year = "1983"
}

@article{PhysRevD.34.1237,
  title = {Phase transition induced by cosmic strings},
  author = {Yajnik, U. A.},
  journal = {Phys. Rev. D},
  volume = {34},
  issue = {4},
  pages = {1237--1240},
  numpages = {0},
  year = {1986},
  month = {Aug},
  publisher = {American Physical Society},
  doi = {10.1103/PhysRevD.34.1237},
  url = {https://link.aps.org/doi/10.1103/PhysRevD.34.1237}
}

@article{Balkin:2021zfd,
    author = "Balkin, Reuven and Serra, Javi and Springmann, Konstantin and Stelzl, Stefan and Weiler, Andreas",
    title = "{Density induced vacuum instability}",
    eprint = "2105.13354",
    archivePrefix = "arXiv",
    primaryClass = "hep-ph",
    reportNumber = "TUM-HEP-1308/20",
    doi = "10.21468/SciPostPhys.14.4.071",
    journal = "SciPost Phys.",
    volume = "14",
    number = "4",
    pages = "071",
    year = "2023"
}

@article{Mukaida:2017bgd,
    author = "Mukaida, Kyohei and Yamada, Masaki",
    title = "{False Vacuum Decay Catalyzed by Black Holes}",
    eprint = "1706.04523",
    archivePrefix = "arXiv",
    primaryClass = "hep-th",
    reportNumber = "IPMU-17-0069",
    doi = "10.1103/PhysRevD.96.103514",
    journal = "Phys. Rev. D",
    volume = "96",
    number = "10",
    pages = "103514",
    year = "2017"
}

@article{Jinno:2023vnr,
    author = "Jinno, Ryusuke and Kume, Jun'ya and Yamada, Masaki",
    title = "{Super-slow phase transition catalyzed by BHs and the birth of baby BHs}",
    eprint = "2310.06901",
    archivePrefix = "arXiv",
    primaryClass = "hep-ph",
    reportNumber = "TU-1209, RESCEU-18/23",
    doi = "10.1016/j.physletb.2024.138465",
    journal = "Phys. Lett. B",
    volume = "849",
    pages = "138465",
    year = "2024"
}

@article{Agrawal:2022hnf,
    author = "Agrawal, Prateek and Nee, Michael",
    title = "{The Boring Monopole}",
    eprint = "2202.11102",
    archivePrefix = "arXiv",
    primaryClass = "hep-ph",
    doi = "10.21468/SciPostPhys.13.3.049",
    journal = "SciPost Phys.",
    volume = "13",
    number = "3",
    pages = "049",
    year = "2022"
}

@article{Jinno:2021ury,
    author = "Jinno, Ryusuke and Konstandin, Thomas and Rubira, Henrique and van de Vis, Jorinde",
    title = "{Effect of density fluctuations on gravitational wave production in first-order phase transitions}",
    eprint = "2108.11947",
    archivePrefix = "arXiv",
    primaryClass = "astro-ph.CO",
    reportNumber = "DESY 21-131, DESY 21-131",
    doi = "10.1088/1475-7516/2021/12/019",
    journal = "JCAP",
    volume = "12",
    number = "12",
    pages = "019",
    year = "2021"
}

@article{Blasi:2023rqi,
    author = "Blasi, Simone and Jinno, Ryusuke and Konstandin, Thomas and Rubira, Henrique and Stomberg, Isak",
    title = "{Gravitational waves from defect-driven phase transitions: domain walls}",
    eprint = "2302.06952",
    archivePrefix = "arXiv",
    primaryClass = "astro-ph.CO",
    doi = "10.1088/1475-7516/2023/10/051",
    journal = "JCAP",
    volume = "10",
    pages = "051",
    year = "2023"
}

@article{Kuzmin:1985mm,
    author = "Kuzmin, V. A. and Rubakov, V. A. and Shaposhnikov, M. E.",
    title = "{On the Anomalous Electroweak Baryon Number Nonconservation in the Early Universe}",
    reportNumber = "IC/85/8",
    doi = "10.1016/0370-2693(85)91028-7",
    journal = "Phys. Lett. B",
    volume = "155",
    pages = "36",
    year = "1985"
}

@article{Azatov:2022tii,
    author = "Azatov, Aleksandr and Barni, Giulio and Chakraborty, Sabyasachi and Vanvlasselaer, Miguel and Yin, Wen",
    title = "{Ultra-relativistic bubbles from the simplest Higgs portal and their cosmological consequences}",
    eprint = "2207.02230",
    archivePrefix = "arXiv",
    primaryClass = "hep-ph",
    reportNumber = "SISSA 12/2022/FISI TU-1157",
    doi = "10.1007/JHEP10(2022)017",
    journal = "JHEP",
    volume = "10",
    pages = "017",
    year = "2022"
}

@article{Azatov:2021irb,
    author = "Azatov, Aleksandr and Vanvlasselaer, Miguel and Yin, Wen",
    title = "{Baryogenesis via relativistic bubble walls}",
    eprint = "2106.14913",
    archivePrefix = "arXiv",
    primaryClass = "hep-ph",
    reportNumber = "SISSA 13/2021/FISI TU-1127",
    doi = "10.1007/JHEP10(2021)043",
    journal = "JHEP",
    volume = "10",
    pages = "043",
    year = "2021"
}

@article{Kurup:2017dzf,
    author = "Kurup, Gowri and Perelstein, Maxim",
    title = "{Dynamics of Electroweak Phase Transition In Singlet-Scalar Extension of the Standard Model}",
    eprint = "1704.03381",
    archivePrefix = "arXiv",
    primaryClass = "hep-ph",
    doi = "10.1103/PhysRevD.96.015036",
    journal = "Phys. Rev. D",
    volume = "96",
    number = "1",
    pages = "015036",
    year = "2017"
}

@inproceedings{Quiros:1999jp,
    author = "Quiros, Mariano",
    title = "{Finite temperature field theory and phase transitions}",
    booktitle = "{ICTP Summer School in High-Energy Physics and Cosmology}",
    eprint = "hep-ph/9901312",
    archivePrefix = "arXiv",
    reportNumber = "IEM-FT-187-99",
    pages = "187--259",
    month = "1",
    year = "1999"
}

@article{Witten:1984rs,
      author         = "Witten, Edward",
      title          = "{Cosmic Separation of Phases}",
      journal        = "Phys. Rev.",
      volume         = "D30",
      year           = "1984",
      pages          = "272-285",
      doi            = "10.1103/PhysRevD.30.272",
      reportNumber   = "PRINT-84-0400 (IAS,PRINCETON)",
      SLACcitation   = "%%CITATION = PHRVA,D30,272;%%"
}

@article{Linde:1981zj,
      author         = "Linde, Andrei D.",
      title          = "{Decay of the False Vacuum at Finite Temperature}",
      journal        = "Nucl. Phys.",
      volume         = "B216",
      year           = "1983",
      pages          = "421",
      doi            = "10.1016/0550-3213(83)90293-6,
                        10.1016/0550-3213(83)90072-X",
      note           = "[Erratum: Nucl. Phys.B223,544(1983)]",
      reportNumber   = "LEBEDEV-81-265",
      SLACcitation   = "%%CITATION = NUPHA,B216,421;%%"
}

@article{Linde:1980tt,
      author         = "Linde, Andrei D.",
      title          = "{Fate of the False Vacuum at Finite Temperature: Theory
                        and Applications}",
      journal        = "Phys. Lett.",
      volume         = "100B",
      year           = "1981",
      pages          = "37-40",
      doi            = "10.1016/0370-2693(81)90281-1",
      reportNumber   = "LEBEDEV-80-92",
      SLACcitation   = "%%CITATION = PHLTA,100B,37;%%"
}

@article{Kosowsky:1992rz,
      author         = "Kosowsky, Arthur and Turner, Michael S. and Watkins,
                        Richard",
      title          = "{Gravitational waves from first order cosmological phase
                        transitions}",
      journal        = "Phys. Rev. Lett.",
      volume         = "69",
      year           = "1992",
      pages          = "2026-2029",
      doi            = "10.1103/PhysRevLett.69.2026",
      reportNumber   = "FERMILAB-PUB-91-333-A-REV, FERMILAB-PUB-91-333-A",
      SLACcitation   = "%%CITATION = PRLTA,69,2026;%%"
}

@article{Kosowsky:1992vn,
      author         = "Kosowsky, Arthur and Turner, Michael S.",
      title          = "{Gravitational radiation from colliding vacuum bubbles:
                        envelope approximation to many bubble collisions}",
      journal        = "Phys. Rev.",
      volume         = "D47",
      year           = "1993",
      pages          = "4372-4391",
      doi            = "10.1103/PhysRevD.47.4372",
      eprint         = "astro-ph/9211004",
      archivePrefix  = "arXiv",
      primaryClass   = "astro-ph",
      reportNumber   = "FERMILAB-PUB-92-295-A",
      SLACcitation   = "%%CITATION = ASTRO-PH/9211004;%%"
}

@article{Kamionkowski:1993fg,
      author         = "Kamionkowski, Marc and Kosowsky, Arthur and Turner,
                        Michael S.",
      title          = "{Gravitational radiation from first order phase
                        transitions}",
      journal        = "Phys. Rev.",
      volume         = "D49",
      year           = "1994",
      pages          = "2837-2851",
      doi            = "10.1103/PhysRevD.49.2837",
      eprint         = "astro-ph/9310044",
      archivePrefix  = "arXiv",
      primaryClass   = "astro-ph",
      reportNumber   = "IASSNS-HEP-93-44, FERMILAB-PUB-93-235-A",
      SLACcitation   = "%%CITATION = ASTRO-PH/9310044;%%"
}

@article{Baldes:2023fsp,
    author = "Baldes, Iason and Dichtl, Maximilian and Gouttenoire, Yann and Sala, Filippo",
    title = "{Bubbletrons}",
    eprint = "2306.15555",
    archivePrefix = "arXiv",
    primaryClass = "hep-ph",
    month = "6",
    year = "2023"
}

@article{Huang:2022vkf,
    author = "Huang, Peisi and Xie, Ke-Pan",
    title = "{Leptogenesis triggered by a first-order phase transition}",
    eprint = "2206.04691",
    archivePrefix = "arXiv",
    primaryClass = "hep-ph",
    doi = "10.1007/JHEP09(2022)052",
    journal = "JHEP",
    volume = "09",
    pages = "052",
    year = "2022"
}

@article{Jung:2021mku,
    author = "Jung, Tae Hyun and Okui, Takemichi",
    title = "{Primordial black holes from bubble collisions during a first-order phase transition}",
    eprint = "2110.04271",
    archivePrefix = "arXiv",
    primaryClass = "hep-ph",
    reportNumber = "KEK-TH-2350",
    month = "10",
    year = "2021"
}

@article{Kawana:2021tde,
    author = "Kawana, Kiyoharu and Xie, Ke-Pan",
    title = "{Primordial black holes from a cosmic phase transition: The collapse of Fermi-balls}",
    eprint = "2106.00111",
    archivePrefix = "arXiv",
    primaryClass = "astro-ph.CO",
    doi = "10.1016/j.physletb.2021.136791",
    journal = "Phys. Lett. B",
    volume = "824",
    pages = "136791",
    year = "2022"
}

@article{Gouttenoire:2023naa,
    author = "Gouttenoire, Yann and Volansky, Tomer",
    title = "{Primordial Black Holes from Supercooled Phase Transitions}",
    eprint = "2305.04942",
    archivePrefix = "arXiv",
    primaryClass = "hep-ph",
    month = "5",
    year = "2023"
}

@article{Lewicki:2023ioy,
    author = "Lewicki, Marek and Toczek, Piotr and Vaskonen, Ville",
    title = "{Primordial black holes from strong first-order phase transitions}",
    eprint = "2305.04924",
    archivePrefix = "arXiv",
    primaryClass = "astro-ph.CO",
    month = "5",
    year = "2023"
}

@article{Bringmann:2023opz,
    author = "Bringmann, Torsten and Depta, Paul Frederik and Konstandin, Thomas and Schmidt-Hoberg, Kai and Tasillo, Carlo",
    title = "{Does NANOGrav observe a dark sector phase transition?}",
    eprint = "2306.09411",
    archivePrefix = "arXiv",
    primaryClass = "astro-ph.CO",
    reportNumber = "DESY-23-077",
    month = "6",
    year = "2023"
}

@article{Enqvist:1991xw,
      author         = "Enqvist, K. and Ignatius, J. and Kajantie, K. and
                        Rummukainen, K.",
      title          = "{Nucleation and bubble growth in a first order
                        cosmological electroweak phase transition}",
      journal        = "Phys. Rev.",
      volume         = "D45",
      year           = "1992",
      pages          = "3415-3428",
      doi            = "10.1103/PhysRevD.45.3415",
      reportNumber   = "HU-TFT-91-35",
      SLACcitation   = "%%CITATION = PHRVA,D45,3415;%%"
}

@article{Long:2017rdo,
    author = "Long, Andrew J. and Tesi, Andrea and Wang, Lian-Tao",
    title = "{Baryogenesis at a Lepton-Number-Breaking Phase Transition}",
    eprint = "1703.04902",
    archivePrefix = "arXiv",
    primaryClass = "hep-ph",
    doi = "10.1007/JHEP10(2017)095",
    journal = "JHEP",
    volume = "10",
    pages = "095",
    year = "2017"
}

@article{Cline:2017jvp,
    author = "Cline, James M.",
    editor = "Auge, Etienne and Dumarchez, Jacques and Tran Thanh Van, Jean",
    title = "{Is electroweak baryogenesis dead?}",
    eprint = "1704.08911",
    archivePrefix = "arXiv",
    primaryClass = "hep-ph",
    reportNumber = "CERN-TH-2017-096",
    doi = "10.1098/rsta.2017.0116",
    journal = "Phil. Trans. Roy. Soc. Lond. A",
    volume = "376",
    number = "2114",
    pages = "20170116",
    year = "2018"
}

@article{Caprini:2011uz,
    author = "Caprini, Chiara and No, Jose M.",
    title = "{Supersonic Electroweak Baryogenesis: Achieving Baryogenesis for Fast Bubble Walls}",
    eprint = "1111.1726",
    archivePrefix = "arXiv",
    primaryClass = "hep-ph",
    reportNumber = "SACLAY-T11-209, ULB-TH-11-25",
    doi = "10.1088/1475-7516/2012/01/031",
    journal = "JCAP",
    volume = "01",
    pages = "031",
    year = "2012"
}

@article{Azatov:2020nbe,
    author = "Azatov, Aleksandr and Vanvlasselaer, Miguel",
    title = "{Phase transitions in perturbative walking dynamics}",
    eprint = "2003.10265",
    archivePrefix = "arXiv",
    primaryClass = "hep-ph",
    reportNumber = "SISSA 03/2020/FISI",
    doi = "10.1007/JHEP09(2020)085",
    journal = "JHEP",
    volume = "09",
    pages = "085",
    year = "2020"
}

@article{Falkowski:2012fb,
    author = "Falkowski, Adam and No, Jose M.",
    title = "{Non-thermal Dark Matter Production from the Electroweak Phase Transition: Multi-TeV WIMPs and 'Baby-Zillas'}",
    eprint = "1211.5615",
    archivePrefix = "arXiv",
    primaryClass = "hep-ph",
    reportNumber = "ULB-TH-12-18, LPT-12-113",
    doi = "10.1007/JHEP02(2013)034",
    journal = "JHEP",
    volume = "02",
    pages = "034",
    year = "2013"
}

@article{Morrissey:2012db,
    author = "Morrissey, David E. and Ramsey-Musolf, Michael J.",
    title = "{Electroweak baryogenesis}",
    eprint = "1206.2942",
    archivePrefix = "arXiv",
    primaryClass = "hep-ph",
    reportNumber = "NPAC-12-08",
    doi = "10.1088/1367-2630/14/12/125003",
    journal = "New J. Phys.",
    volume = "14",
    pages = "125003",
    year = "2012"
}

@article{Craig:2020jfv,
    author = "Craig, Nathaniel and Levi, Noam and Mariotti, Alberto and Redigolo, Diego",
    title = "{Ripples in Spacetime from Broken Supersymmetry}",
    eprint = "2011.13949",
    archivePrefix = "arXiv",
    primaryClass = "hep-ph",
    doi = "10.1007/JHEP02(2021)184",
    journal = "JHEP",
    volume = "21",
    pages = "184",
    year = "2020"
}

@article{vonHarling:2019gme,
    author = "Von Harling, Benedict and Pomarol, Alex and Pujol\`as, Oriol and Rompineve, Fabrizio",
    title = "{Peccei-Quinn Phase Transition at LIGO}",
    eprint = "1912.07587",
    archivePrefix = "arXiv",
    primaryClass = "hep-ph",
    doi = "10.1007/JHEP04(2020)195",
    journal = "JHEP",
    volume = "04",
    pages = "195",
    year = "2020"
}

@article{ET:2019dnz,
    author = "Maggiore, Michele and others",
    collaboration = "ET",
    title = "{Science Case for the Einstein Telescope}",
    eprint = "1912.02622",
    archivePrefix = "arXiv",
    primaryClass = "astro-ph.CO",
    doi = "10.1088/1475-7516/2020/03/050",
    journal = "JCAP",
    volume = "03",
    pages = "050",
    year = "2020"
}

@article{Romero:2021kby,
    author = "Romero, Alba and Martinovic, Katarina and Callister, Thomas A. and Guo, Huai-Ke and Mart\'\i{}nez, Mario and Sakellariadou, Mairi and Yang, Feng-Wei and Zhao, Yue",
    title = "{Implications for First-Order Cosmological Phase Transitions from the Third LIGO-Virgo Observing Run}",
    eprint = "2102.01714",
    archivePrefix = "arXiv",
    primaryClass = "hep-ph",
    doi = "10.1103/PhysRevLett.126.151301",
    journal = "Phys. Rev. Lett.",
    volume = "126",
    number = "15",
    pages = "151301",
    year = "2021"
}

@article{Cutting:2022zgd,
    author = "Cutting, Daniel and Vilhonen, Essi and Weir, David J.",
    title = "{Droplet collapse during strongly supercooled transitions}",
    eprint = "2204.03396",
    archivePrefix = "arXiv",
    primaryClass = "astro-ph.CO",
    reportNumber = "HIP-2022-5/TH",
    doi = "10.1103/PhysRevD.106.103524",
    journal = "Phys. Rev. D",
    volume = "106",
    number = "10",
    pages = "103524",
    year = "2022"
}

@article{Bruggisser:2022rdm,
    author = "Bruggisser, Sebastian and von Harling, Benedict and Matsedonskyi, Oleksii and Servant, Geraldine",
    title = "{Status of electroweak baryogenesis in minimal composite Higgs}",
    eprint = "2212.11953",
    archivePrefix = "arXiv",
    primaryClass = "hep-ph",
    reportNumber = "DESY-22-209",
    doi = "10.1007/JHEP08(2023)012",
    journal = "JHEP",
    volume = "08",
    pages = "012",
    year = "2023"
}

@article{Leitao:2014pda,
    author = "Leitao, Leonardo and Megevand, Ariel",
    title = "{Hydrodynamics of phase transition fronts and the speed of sound in the plasma}",
    eprint = "1410.3875",
    archivePrefix = "arXiv",
    primaryClass = "hep-ph",
    doi = "10.1016/j.nuclphysb.2014.12.008",
    journal = "Nucl. Phys. B",
    volume = "891",
    pages = "159--199",
    year = "2015"
}

@article{Azatov:2024crd,
    author = "Azatov, Aleksandr and Nagels, Xander and Vanvlasselaer, Miguel and Yin, Wen",
    title = "{Populating secluded dark sector with ultra-relativistic bubbles}",
    eprint = "2406.12554",
    archivePrefix = "arXiv",
    primaryClass = "hep-ph",
    reportNumber = "SISSA 11/2024/FISI",
    doi = "10.1007/JHEP11(2024)129",
    journal = "JHEP",
    volume = "11",
    pages = "129",
    year = "2024"
}

@article{Katz:2009gh,
    author = "Katz, Andrey",
    title = "{On the Thermal History of Calculable Gauge Mediation}",
    eprint = "0907.3930",
    archivePrefix = "arXiv",
    primaryClass = "hep-th",
    reportNumber = "UMD-PP-09-045",
    doi = "10.1088/1126-6708/2009/10/054",
    journal = "JHEP",
    volume = "10",
    pages = "054",
    year = "2009"
}

@article{Craig:2006kx,
    author = "Craig, Nathaniel J. and Fox, Patrick J. and Wacker, Jay G.",
    title = "{Reheating Metastable O'Raifeartaigh Models}",
    eprint = "hep-th/0611006",
    archivePrefix = "arXiv",
    reportNumber = "SLAC-PUB-12215",
    doi = "10.1103/PhysRevD.75.085006",
    journal = "Phys. Rev. D",
    volume = "75",
    pages = "085006",
    year = "2007"
}

@article{ORaifeartaigh:1975nky,
    author = "O'Raifeartaigh, L.",
    title = "{Spontaneous Symmetry Breaking for Chiral Scalar Superfields}",
    reportNumber = "DIAS-TP-75-9",
    doi = "10.1016/0550-3213(75)90585-4",
    journal = "Nucl. Phys. B",
    volume = "96",
    pages = "331--352",
    year = "1975"
}

@article{Intriligator:2007py,
    author = "Intriligator, Kenneth A. and Seiberg, Nathan and Shih, David",
    title = "{Supersymmetry breaking, R-symmetry breaking and metastable vacua}",
    eprint = "hep-th/0703281",
    archivePrefix = "arXiv",
    reportNumber = "UCSD-PTH-06-12",
    doi = "10.1088/1126-6708/2007/07/017",
    journal = "JHEP",
    volume = "07",
    pages = "017",
    year = "2007"
}

@article{Nelson:1993nf,
    author = "Nelson, Ann E. and Seiberg, Nathan",
    title = "{R symmetry breaking versus supersymmetry breaking}",
    eprint = "hep-ph/9309299",
    archivePrefix = "arXiv",
    reportNumber = "UCSD-PTH-93-27, RU-93-42",
    doi = "10.1016/0550-3213(94)90577-0",
    journal = "Nucl. Phys. B",
    volume = "416",
    pages = "46--62",
    year = "1994"
}

@article{Craig:2009zx,
    author = "Craig, Nathaniel J.",
    title = "{Gravitational Waves from Supersymmetry Breaking}",
    eprint = "0902.1990",
    archivePrefix = "arXiv",
    primaryClass = "hep-ph",
    reportNumber = "SU-ITP-09-08",
    month = "2",
    year = "2009"
}

@article{Ai:2024uyw,
    author = "Ai, Wen-Yuan and Laurent, Benoit and van de Vis, Jorinde",
    title = "{Bounds on the bubble wall velocity}",
    eprint = "2411.13641",
    archivePrefix = "arXiv",
    primaryClass = "hep-ph",
    reportNumber = "CERN-TH-2024-198, KCL-PH-TH/2024-57",
    month = "11",
    year = "2024"
}

@article{PhysRevD.7.1888,
  title = {Radiative Corrections as the Origin of Spontaneous Symmetry Breaking},
  author = {Coleman, Sidney and Weinberg, Erick},
  journal = {Phys. Rev. D},
  volume = {7},
  issue = {6},
  pages = {1888--1910},
  numpages = {0},
  year = {1973},
  month = {Mar},
  publisher = {American Physical Society},
  doi = {10.1103/PhysRevD.7.1888},
  url = {https://link.aps.org/doi/10.1103/PhysRevD.7.1888}
}

@article{Barni:2024lkj,
    author = "Barni, Giulio and Blasi, Simone and Vanvlasselaer, Miguel",
    title = "{The hydrodynamics of inverse phase transitions}",
    eprint = "2406.01596",
    archivePrefix = "arXiv",
    primaryClass = "hep-ph",
    doi = "10.1088/1475-7516/2024/10/042",
    journal = "JCAP",
    volume = "10",
    pages = "042",
    year = "2024"
}

@article{Bea:2024bls,
    author = "Bea, Yago and Giliberti, Mauro and Mateos, David and Sanchez-Garitaonandia, Mikel and Serantes, Alexandre and Zilh\~ao, Miguel",
    title = "{Bubble dynamics in a QCD-like phase diagram}",
    eprint = "2412.09588",
    archivePrefix = "arXiv",
    primaryClass = "hep-th",
    month = "12",
    year = "2024"
}

@article{Bea:2024bxu,
    author = "Bea, Yago and Casalderrey-Solana, Jorge and Mateos, David and Sanchez-Garitaonandia, Mikel",
    title = "{Hydrodynamics of Relativistic Superheated Bubbles}",
    eprint = "2406.14450",
    archivePrefix = "arXiv",
    primaryClass = "hep-th",
    reportNumber = "CPHT-RR050.062024",
    month = "6",
    year = "2024"
}

@article{Balkin:2021wea,
    author = "Balkin, Reuven and Serra, Javi and Springmann, Konstantin and Stelzl, Stefan and Weiler, Andreas",
    title = "{Runaway relaxion from finite density}",
    eprint = "2106.11320",
    archivePrefix = "arXiv",
    primaryClass = "hep-ph",
    reportNumber = "TUM-HEP-1348/21",
    doi = "10.1007/JHEP06(2022)023",
    journal = "JHEP",
    volume = "06",
    pages = "023",
    year = "2022"
}

@article{Balkin:2023xtr,
    author = "Balkin, Reuven and Serra, Javi and Springmann, Konstantin and Stelzl, Stefan and Weiler, Andreas",
    title = "{Heavy neutron stars from light scalars}",
    eprint = "2307.14418",
    archivePrefix = "arXiv",
    primaryClass = "hep-ph",
    reportNumber = "IFT-UAM/CSIC-23-96 and TUM-HEP-1468/23",
    doi = "10.1007/JHEP02(2025)141",
    journal = "JHEP",
    volume = "02",
    pages = "141",
    year = "2025"
}
\bibliographystyle{JHEP}

\begin{appendix}

\begin{widetext}

\section{Inverse FOPTs in the $\mu\nu$ model}
\label{app:munu}

In this section, we examine the emergence of inverse phase transitions in the \( \mu\nu \)-model~\cite{Leitao:2014pda}, also referred to as the \( \nu \)-model in Ref.~\cite{Giese:2020rtr} and the template model in Refs.~\cite{Giese:2020znk,Ai:2023see,Sanchez-Garitaonandia:2023zqz}. The \( \mu\nu \)-model extends the standard bag model by allowing the sound speed to deviate from the relativistic value of \( 1/\sqrt{3} \), while remaining constant within each phase.
Explicitly, the EoS for the symmetric and broken phases is given by
\begin{align}
\label{eq:nu_eos}
    &e_\pm(T)= a_\pm T^{\nu_\pm}+\epsilon_\pm\,,\quad\qquad \ \ p_\pm(T)=c_{s,+}^2a_\pm T^{\nu_\pm}-\epsilon_\pm\,, \quad\qquad \nu_\pm=1+1/{c^2_{s,\pm}} \, . 
\end{align}
In the following $\nu_+\equiv \nu$ and $\nu_-\equiv\mu$, we consider $\mu > \nu$ as this mimics the thermal history of the $R$--symmetry model, presented in the main text.
The velocity relations from the matching conditions take the form
\begin{subequations}
\begin{align}
v_+ v_- = \frac{\mu - \mu\nu -r\nu(3 \alpha_\theta-1)(\mu-1)}
{\big(\mu - \mu \nu + r\nu(3 \alpha_\theta +\mu -1)\big)(\mu-1)} \, , \qquad
\frac{v_+}{v_-} = \frac{(\mu-1)\big(\mu - \mu\nu+r\nu(3 \alpha_\theta-1)\big)}
{\mu - \mu\nu- r\nu(3 \alpha_\theta + \mu-1)(\mu-1)} \, ,
\end{align}
\end{subequations}
where we define the ratio \(  r \equiv {a_+ T^\nu_+}/{a_- T^\mu_-} \). Additionally, the strength parameter \( \alpha_\theta \), defined from the pseudo-trace $\theta$ as $\alpha_\theta \equiv {4 D\theta}/{3 w_+}$ where $\theta= e-p/{c_{s,-}^2}$,
within the $\mu\nu$ model evaluates to
\begin{equation}
\label{eq:alphaL}
\alpha_\theta =  \frac{\nu - 1}{3\nu} \left( \frac{\nu - \mu}{\nu - 1} + \mu \alpha_+ \right)\,,
 \qquad \alpha_+ \equiv  \frac{\Delta \epsilon}{a_+ T^\nu_+}=\frac{\epsilon_+-\epsilon_-}{a_+ T^\nu_+}  \, . 
\end{equation}
It is important to emphasize that \( \alpha_\theta \) serves as the fundamental quantity determining the nature of the transition, and directly corresponds to the strength of the phase transition computed via the pseudo-trace.

Notably, in the case of the traditional bag EoS, where \( \mu=\nu=4 \), the pseudo-trace coincides with the standard definition of the phase transition strength, \( \alpha_\theta  = \alpha_+ \), thereby recovering the standard velocity relations.

It is shown in Fig.~\ref{fig:branches} that, as soon as \( \alpha_\theta < 0 \), the \textit{inverse branches} emerge. This confirms that in the \( \mu\nu \)-model, a negative \( \alpha_\theta \) implies an inverse phase transition. Analogously, for the bag EoS, a negative \( \alpha_+ \) corresponds to an inverse PT. This result aligns with the characterization proposed in~\cite{Barni:2024lkj}, where it was shown that within the bag EoS, \( \Delta \epsilon < 0 \) serves as a direct indicator of an inverse phase transition.

\begin{figure}
    \centering
    \includegraphics[width=0.5\linewidth]{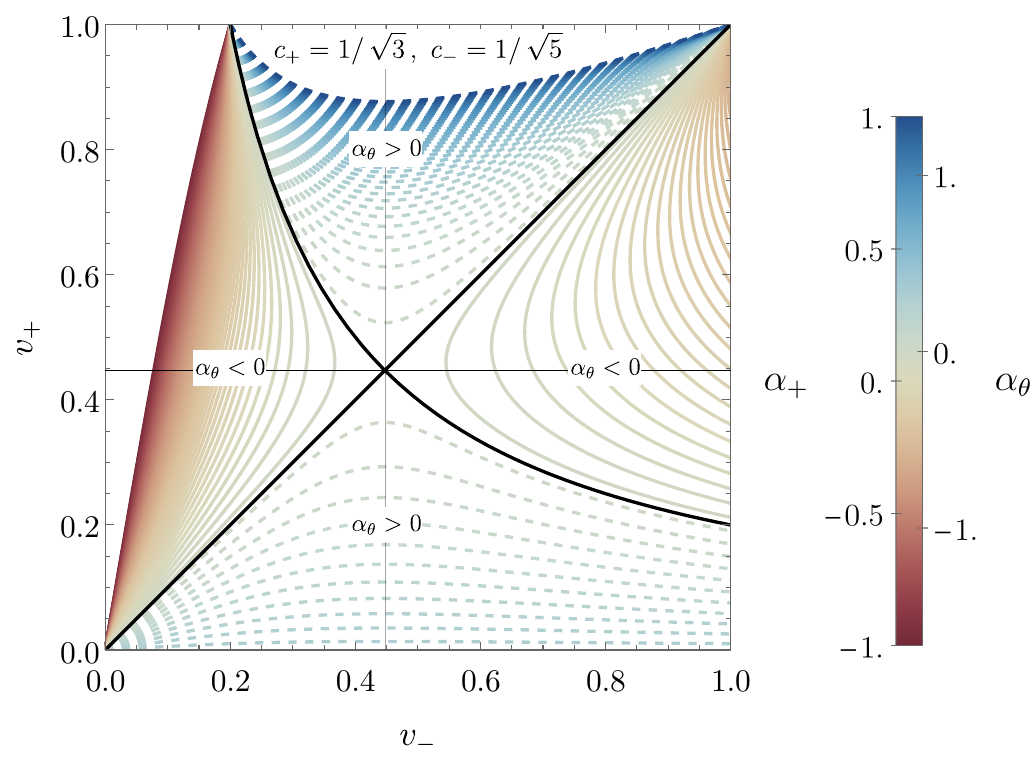}
    \caption{Dashed (solid) lines represent direct (inverse) phase transitions. The inverse branches emerge as soon as $\alpha_\theta < 0$, whereas this is not necessarily the case for $\alpha_+$. The two strength parameters of the phase transition, $\alpha_+$ and $\alpha_\theta$, coincide in the bag model when $\mu = \nu = 4$. 
    }
    \label{fig:branches}
\end{figure}

\section{Solving the Hydrodynamic Equations for the Fluid Profiles}
\label{app:hydro}
The conservation of the energy-momentum tensor for a relativistic fluid, given by $\nabla_\mu T^{\mu\nu} = 0$,
yields two independent hydrodynamic equations. These equations can be rewritten in terms of the enthalpy density, \( w = e + p \). We consider a spherically symmetric and self-similar configuration, where the fluid variables depend only on \( \xi \equiv r/t \), the similarity variable. Using this variable, it can be shown that the hydrodynamics equations, in terms of the fluid velocity \( v(\xi) \) and the fluid temperature \( T(\xi) \), take the following form
\begin{align}
\label{eq: hydro_system}
    (\xi - v) \frac{\partial_\xi T}{w}\frac{de}{dT} &= \frac{2 v}{\xi} + [1 - \gamma^2 v (\xi - v)] \partial_\xi v\,, \qquad
    \frac{\partial_\xi T}{T} = \gamma^2 \mu(\xi, v) \partial_\xi v\,,
\end{align}
where $\mu(\xi, v)=\frac{ \xi - v}{1 - \xi v}$.
It is important to emphasize that the thermodynamic quantities, such as \( p \) and \( e \), must be evaluated in the appropriate phase depending on the region where the equation is being solved. In the remainder of this section, we present the different types of expansion modes for inverse PTs within this general framework.

\paragraph{\textbf{Inverse Deflagration}}  
To fully specify the system of equations in Eqs.~\eqref{eq: hydro_system}, we must define the initial conditions for \( v(\xi) \) and \( T(\xi) \). In the case of an inverse deflagration, this translates to  
\begin{align}
    \xi_w = v_+\,, \qquad v(\xi_w) = \mu(v_+, v_-)\,,\qquad
    T(\xi_w^+) = T_+\,, \qquad T(\xi_w^-) = T_-\,,
\end{align}
where the \( + \) phase corresponds to the false vacuum, while the \( - \) phase corresponds to the true vacuum.
Additionally, we impose the condition for the formation of a shock wave, which is given by  
$
    \mu(\xi_{ sh}, v(\xi_{ sh})) \xi_{sh} = c_{s,-}^2(T(\xi_{ sh})) \, .
$
These initial conditions also apply to standard detonations, provided that the pair \( (v_+, v_-) \) satisfies the condition \( v_+ > v_- \). 

\paragraph{\textbf{Inverse Detonations}}  
For inverse detonations, the initial conditions across the discontinuity translate into
\begin{align}
    \xi_w = v_-\,, \qquad v(\xi_w) = \mu(v_-, v_+)\,, \qquad
    T(\xi_w^+) = T_+\,, \qquad T(\xi_w^-) = T_- \, .
\end{align}
It can be checked directly that the rarefaction wave terminates at $
    \xi_{\rm end} = c_{s,+}(T(\xi_{\rm end}))$.
For a standard detonation, the substitution \( c_{s,+} \to c_{s,-} \) must be applied, as the rarefaction wave develops behind the reaction front, i.e., in the new phase.

These initial conditions also apply to standard deflagrations, provided that the pair \( (v_+, v_-) \) satisfies the appropriate conditions. In this case, the shock condition must be modified by replacing \( c_{s,-} \) with \( c_{s,+} \), as the shock forms ahead of the reaction front in the old phase.

Before discussing the last type of solution, it is important to highlight the presence of strong solutions in Fig.~\ref{fig:crossing}, where the red-shaded region indicates their domain. For (inverse) detonations/deflagrations, the strong regime is defined by the conditions $(v_+ \gtrless c_{s,+}(T_+))\ v_-\lessgtr c_{s,-}(T_-)$. As previously discussed in~\cite{Barni:2024lkj}, strong (inverse) detonations cannot be consistently realised, while strong (inverse) deflagrations, although they may initially form due to the dynamics of the phase transition, are inherently unstable. Over time, they will decay into (inverse) hybrid solutions.

\paragraph{\textbf{Inverse Hybrid}}  
For inverse hybrid solutions, as in the standard case, to make the profile stable, we must connect a strong inverse deflagration to a Chapman-Jouguet inverse detonation, which is defined as a detonation with \( v_+ = c_{s,+}(T_+) \). The initial conditions then translate into
\begin{align}
    v(\xi_w^+) = \mu(\xi_w^+, c_{s,+}(T_+))\,, \qquad v(\xi_w^-) = \mu(\xi_w^-, v_-)\,, \qquad
    T(\xi_w^+) = T_+\,, \qquad T(\xi_w^-) = T_- \, ,
\end{align}
where the four input parameters required to specify the system are \( (\xi_w, v_-, T_+, T_-) \). 

Additionally, the shock formation condition must be imposed, and one can verify that the rarefaction wave of the inverse detonation terminates again at $
    \xi_{\rm end} = c_{s,+}(T(\xi_{\rm end})) \, .$ The maximal range of wall velocities for which an inverse hybrid solution exists is given by $c_{s,-}^2 < \xi_w < c_{s,+}$ where the lower bound arises because the slowest possible inverse hybrid is determined by the slowest possible shock.

For the case of a direct hybrid transition, a strong deflagration must instead be connected to a CJ detonation, where the latter is characterized by \( v_- = c_{s,-}(T_-) \). The allowed range of wall velocities in this case is $c_{s,-} < \xi_w < 1$ where the upper bound is simply the speed of light, as there is no fundamental constraint on the maximum speed of the shock front.

\paragraph{\textbf{Overlap in the hybrid corner}}
\begin{figure}
    \centering
    \includegraphics[width=0.318\linewidth]{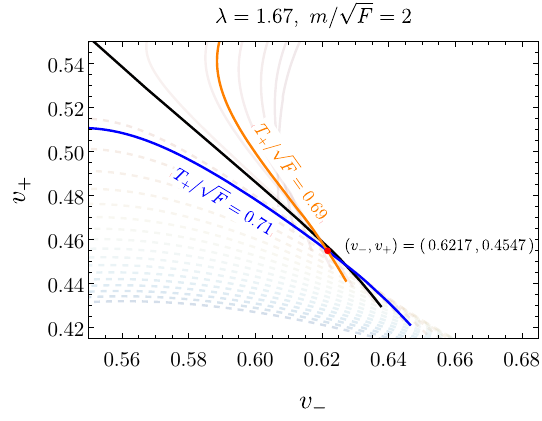} \includegraphics[width=0.327\linewidth]{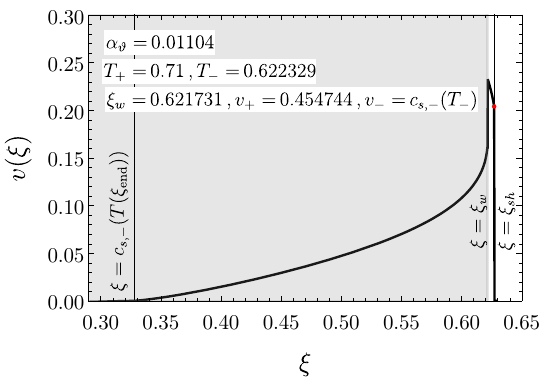} \includegraphics[width=0.327\linewidth]{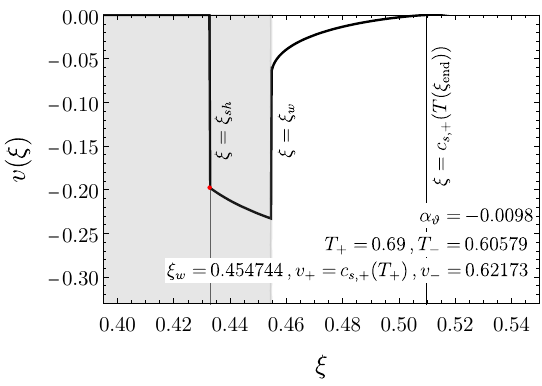}
    \caption{Overlap of direct and inverse branches in the $(v_-, v_+)$ plane and corresponding fluid profiles. \textit{Left panel}: The $(v_-, v_+)$ trajectories for different values of $T_+$. The inverse branch is shown in orange, while the direct branch is displayed in blue. The highlighted crossing point indicates a case where both a direct and an inverse solution exist for the same $(v_-, v_+)$ pair. \textit{Middle panel}: Fluid profile corresponding to the direct hybrid solution. \textit{Right panel}: Fluid profile for the inverse hybrid solution. The shaded regions indicate the interior of the bubble. }
    \label{fig:crossing}
\end{figure}
In our numerical analysis, we observe that in the hybrid transition regime, the branches in the \( (v_-,v_+) \) plane exhibit an overlap between direct and inverse transitions. This is particularly evident when zooming in on the hybrid region, as shown in Fig.~\ref{fig:crossing} (left panel). There, we explicitly construct two distinct solutions corresponding to the same pair of values \( (v_-, v_+) \), demonstrating the existence of overlapping branches, in the middle and right panel of Fig.~\ref{fig:crossing}.

This overlap arises due to the stability conditions required for hybrid solutions. Specifically, for both direct and inverse hybrids to remain stable, the fluid velocity just behind (or in front of) the wall must match the local speed of sound in the respective phase at the corresponding temperature. That is, stability demands that for (inverse) hybrid holds $(v_+ = c_{s,+}(T_+))\ v_- = c_{s,-}(T_-)$. This condition provides additional flexibility in setting \( \xi_w = v_- \) for direct hybrids and \( \xi_w = v_+ \) for inverse hybrids, thus allowing both solutions to coexist.

Another key reason for this overlap is related to the structure of the separatrices (black solid lines) in the \( (v_-, v_+) \) plane. Ideally, these separatrices would be given by $v_- = v_+$ and $v_- v_+ = c_{s,-}^2$, however, since the speed of sound varies along the branches due to temperature dependence, the boundary between the direct and inverse solutions is no longer sharply defined. Despite their overlap in the \( (v_-, v_+) \) plane, the two solutions can still be distinguished physically. Each branch corresponds to a different set of temperatures \( (T_+, T_-) \), leading to a different transition strength characterized by the generalised pseudotrace, \( \alpha_\vartheta \), which will have in fact a different sign. Thus, even though the solutions may appear degenerate in velocity space, they remain distinct due to their thermodynamic properties.

\section{Effective potential}
\label{app:effect_pot}
In this appendix, we outline the computation of the one-loop and thermal corrections to the potential, as used in the main text.

\paragraph{One-Loop Potential.}
It is well known that quantum corrections at one loop modify the shape of the scalar potential~\cite{PhysRevD.7.1888}. At one loop, the tree-level potential is corrected by the \emph{Coleman-Weinberg} potential, given by
\begin{equation}
V_{\rm CW}(x) = \sum_{i=f, s} \frac{n_i (-1)^F}{64\pi^2} 
\left[M_i^4(x) \left(\log \frac{M_i^2(x)}{\Lambda^2} - \frac{3}{2}\right)\right] \, ,
\label{eq:CW_cur_off_reg}
\end{equation}
such that the total \emph{effective potential} is
\begin{equation}
V_0(x) = V_{\rm tree}(x) + V_{\rm CW}(x) \, .
\end{equation}
Here, \( M_i(x) \) denotes the field-dependent masses of the particles in the spectrum, including the four fermionic and four scalar degrees of freedom. The sum runs over all states, with \( F = 1(0) \) for fermions (scalars), and \( \Lambda \) is the renormalization scale, which we set to \( \Lambda = m \).

\paragraph{Thermal corrections.}
In the early universe, high temperatures and the associated thermal fluctuations modify the effective potential. These thermal effects can be incorporated by adding finite-temperature corrections to the zero-temperature potential~\cite{Quiros:1999jp,Curtin:2016urg}, leading to 
\begin{equation}
V_{\rm eff}(T,x) = V_0(x) + V_T(x) \, .
\label{eq:thermal_pot1}
\end{equation}
Here, \( V_0(x) \) is the one--loop effective potential derived above, while the thermal potential \( V_T(x) \) is given by
\begin{align}
V_T(M_i(x)) &= \sum_{i \in B} \frac{n_i}{2\pi^2} T^4 J_B\bigg(\frac{M_i^2(x)}{T^2}\bigg) 
- \sum_{i \in F} \frac{n_i}{2\pi^2} T^4 J_F\bigg(\frac{M_i^2(x)}{T^2}\bigg), \notag
\\ 
J_{B/F}(y^2) &= \int_0^{\infty} dx\ x^2\log \Big[1\mp\exp{(-\sqrt{x^2+y^2})}\Big] \,,
\label{eq:thermal_pot}
\end{align}
where the sum includes the (tree--level) massless fields in the $X$ superfield corresponding to the goldstino, the pseudomodulus, and the $R$--axion, which will give a constant, namely $x$--independent, contribution to the free energy at one loop.

The full one-loop potential, including both quantum and thermal corrections, then takes the standard form
\begin{equation}
V_{\rm eff}(x, T)= V_{\rm tree}(x) + 
\sum_i \left[ V_{\rm CW}\big(M_i^2(x)\big) + V_T\big(M_i^2(x)\big) \right] \, . 
\end{equation}
where we have neglected the thermal masses. In fact, perturbation theory for thermal field theory breaks down for massless particles. This breakdown can be avoided by the resummation of the so-called daisy diagrams, which amounts to add thermal masses to the tree-level ones in the CW and thermal potential. However, as our model does not contain dynamical massless particles on either phase, we expect
this to be a good approximation.
A detailed study of additional thermal effects is left for future work.

\section{Nucleation theory}
\label{app:dynamics}

\begin{figure}
 \centering  \includegraphics[width=0.48\textwidth]{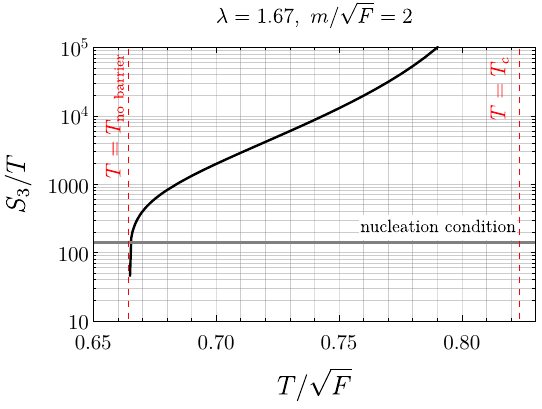}
 \caption{We present the Euclidean action as a function of temperature for the benchmark point under consideration. Assuming a nucleation criterion of \( S_3/T \sim 140 \), this corresponds to setting \( \sqrt{F} \sim \text{TeV} \). As we can see, tunnelling occurs around \( T_{\rm nuc}/\sqrt{F} \sim 0.665 \).}
 \label{fig:Endo}
 \end{figure}
In this section we study the tunnelling rate for the $R$--symmetry breaking FOPT of interest. 
Starting with the effective potential derived in the previous section, we can proceed to the analysis of the phase transition. FOPTs occur when the minima of the effective potential corresponding to different phases are separated by a potential barrier, so that the transition proceeds via the nucleation of bubbles. The probability of bubble nucleation per unit time and unit volume is given by~\cite{Coleman:1977py,Linde:1980tt,Linde:1981zj}
\bea 
\Gamma(T) \simeq \Gamma_3+\Gamma_4= T^4\bigg(\frac{S_3}{2\pi T}\bigg)^{3/2} e^{-S_3(T)/T}+\frac{1}{R_0^4}\bigg(\frac{S_4}{2\pi}\bigg)e^{-S_4}\;,
\label{eq:nucleation_rate}
\eea 
where $S_3$ and $S_4$ are the $O(3)$ and $O(4)$ bounce actions, respectively, and $R_0$ is the bubble radius at nucleation. In the case of interest, the tunnelling is dominantly induced by thermal fluctuations, and we can thus neglect the $O(4)$ contribution. The probability of finding a specific point of the universe in the false vacuum at a given temperature is given by~\cite{Ellis:2019oqb,PhysRevLett.44.963.2}
\bea 
P_f(T) = \exp [-I(T)]\;, \qquad I(T) \equiv \frac{4\pi}{3} \int^{T_c}_T \frac{dT_1\Gamma(T_1) v_w^3}{T_1^4 H(T_1)} \bigg[\int_T^{T_1} \frac{dT_2}{H(T_2)}\bigg]^3.
\label{eq:nucleation}
\eea 
In Eq.(\ref{eq:nucleation_rate}), the strongest dependence on the temperature comes from $\Gamma(T)\propto \exp{(-S_3/T)}$, so that the quantity $I(T)$ is mostly controlled by the ratio $\Gamma\left(T\right)/H\left(T\right)^4$, and one can estimate that, on average, one bubble has nucleated in one Hubble volume when $\Gamma(T)\sim H(T)^4$. The temperature that satisfies this condition is
referred to as the nucleation temperature, $T_{\rm nuc}$. The nucleation condition $\Gamma\sim H^4$ approximately reads
\begin{equation}
    \frac{S_3}{T} \bigg|_{T=T_{\rm nuc}} \sim 4 \log\left(\frac{T_{\rm nuc}}{ H}\right) \sim 140,
\end{equation}
where in the last step we have considered a FOPT occurring around $T_\text{nuc} \sim \ 1\,\text{TeV}$.
One can additionally define the percolation temperature, $T_{\rm per}$, as the temperature when a significant fraction of space, customarily taken to be $\sim$ 34\%, has been converted to the true vacuum:
\bea 
I(T \equiv T_{\rm per}) = 0.34.
\label{eq:condi_per}
\eea 
For relatively fast FOPTs, one however has $T_\text{per} \simeq T_\text{nuc}$.

The bounce solution and the corresponding bounce action are obtained via the well-known overshoot/undershoot method to solve the equations of motion for bubble nucleation. We present the value of the ratio $S_3/T$ as a function of $T$ in Fig.\,\ref{fig:Endo} for the benchmark point with $\lambda = 1.67, m /\sqrt{F}=2$ as in the main text. 

Another important quantity characterising the FOPT is its duration, which is related to the radius of bubbles at collision, $R_\star$, by the approximate relation~\cite{Enqvist:1991xw}:
\begin{align}
  &{\beta \over H} \simeq \frac{(8\pi)^{1/3}}{R_\star H}\;,
 \label{eq:beta2}
 \end{align}
 where $\beta$ is given by
 \bea \beta \equiv -\frac{d}{dt} \frac{S_3}{T}\bigg|_{T = T_{\rm nuc}} = H T \frac{d}{dT}\frac{S_3}{T}\bigg|_{T = T_{\rm nuc}}.
 \eea
For the FOPT under consideration, we find typical values of $\beta/H = \mathcal{O}(10^4)$ in the relevant parameter space. 
Given the short duration of the phase transition, the corresponding GW signal will be suppressed. We leave a comprehensive study of the model parameter space in view of the detectability of current and future GW experiments for future work.

\section{Velocity of the bubble}
\label{app:velo}

Together with the strength, $\alpha_\vartheta$, the duration, $\beta^{-1}$, and the nucleation temperature, another crucial parameter for the description of the phase transition is the velocity of the bubble wall. In principle, there are two qualitatively different possibilities: 1) the bubble wall reaches a steady state, described by an (inverse) deflagration, (inverse) detonation or (inverse) hybrid or the bubble wall keeps accelerating until collision. The following study aims at clarifying which of the two is realised, following the methods presented in~\cite{Ai:2024shx, Barni:2024lkj, Ai:2024uyw}.

\subsection{Collisionless regime computation}
  In principle, the possibility of runaway can be studied in the collisionless limit (see however~\cite{Ai:2024shx}), since in this case, the wall boost factor becomes very large $\gamma_w \gg 1$, the pressure from the exchange of momentum originates from some particles losing their mass and inducing a kick on the wall. 
In the fast wall limit, no particle can escape the bubble, so we can consider only the entering species. To obtain the exchange of momentum, in the wall frame, we can apply the conservation of energy along the particle trajectory,
\bea 
\label{Eq:kick_inverse}
E_i = \sqrt{m_i^2 + p^2_{z,i} + p_\perp^2}\, , \qquad \frac{dE}{dz} =  \bigg(\frac{dm_i^2}{dz} + \frac{dp_{z,i}^2}{dz}\bigg) \frac{1}{2E} = 0\, ,
\qquad  
\Rightarrow \Delta p^{\rm part}_{z,i} \approx -\frac{\Delta m_i^2}{2 p_z}\, ,
\eea   
where $\Delta m^2$ has to be understood as the change of mass of the particle $i$ upon crossing the wall. By conservation of momentum, the wall receives an equal and opposite kick, $ \Delta p^{\rm part}_z = -  \Delta p^{\rm wall}_z > 0 $, which accelerates it forward or backwards depending on the sign of the kick. We observe that a particle \emph{gaining} mass induces a negative kick, and so resists the expansion of the wall, while a particle \emph{losing} mass aspires the wall. In the model under consideration, both types of particles are present, so the competition between them will determine the sign of the collisionless pressure. To capture the pressure induced by the plasma, we need to further convolute the momentum with the incoming flux
\begin{align}
\mathcal{P}_{\rm plasma}
&= \int dz \partial_z  \phi \sum_i g_i\frac{d m^2_i(\phi)}{d\phi}\int \frac{d^3{\bf p}}{(2\pi)^32E_i}\,f_i(p,z, T) 
\approx  \sum_{i}g_i\int \frac{d^3 {\bf p}}{(2\pi)^3}  \frac{\Delta m^2_i}{2 E_i} f^{\rm eq}_{\rm outside}(p, T)  \, . 
\label{P_plasma}
\end{align}
where the sum is to be performed over all particles which may lose or gain a mass across the bubble wall, and $g_i$ is the number of dofs for each particle. By convention, a negative pressure aspires the wall while a positive one resists the expansion. The integral over the phase space is frame-independent and we compute it in the plasma frame. 
 On the other hand, the force exerted by the vacuum energy is given by 
\bea 
F_{\rm vacuum} \equiv \int dz \partial_z \phi \frac{d V^{T=0}(\phi)}{d\phi},  
\eea 
Therefore, evaluating the pressure from Eq.\eqref{P_plasma} in the limit $\gamma_w \to \infty$ if the following inequality is satisfied,
\bea 
\label{eq:runawayinverse}
 \mathcal{P}^{\gamma_w \to \infty}_{\rm plasma} - F_{\rm vacuum}   < 0   \, ,
\eea
then the wall can in principle runaway. Numerical evaluation shows that for the inverse PT window studied in this paper, $ F_{\rm bubble} > 0$, implying that runaway is not possible.

\subsection{Local thermal equilibrium approach}

In the regime of very small velocities, one can approximate that the fluid inside the bubble wall can reach thermalisation, i.e. local thermal equilibrium~\cite{Ai:2021kak,Ai:2023see, Ai:2024shx}. In this case, the entropy current is conserved inside the wall as well and one can solve exactly the matching conditions. In this case, the pushing plasma effect is given by 
\begin{align}
\label{eq:Pfriction_as_a_sum}
    \mathcal{P}_{\rm plasma} \equiv -\int dz \partial_z  \phi \sum_i\frac{d m^2_i(\phi)}{d\phi}\int \frac{d^3{\bf p}}{(2\pi)^32E_i}\,f_i(p,z, T) =\mathcal{P}_{\rm LTE}+\mathcal{P}_{\rm dissipative}  \ .
\end{align}

In the LTE approach, we ignore the dissipative contributions. The approximate LTE expression becomes  (when we can approximate $T_+ \approx T_-$):
\bea 
\label{eq:pressure_budget_inv}
F_{\rm vacuum} - \mathcal{P}_{\rm plasma}  \approx \frac{3w_+}{4}\bigg( \frac{1}{4}(b-1) - |\alpha_+|\bigg)\ ,  \quad b \equiv a_-/a_+ \, , 
\eea 
where now $b> 1$. We observe that the driving force fuelling the expansion now originates from the change of d.o.f. and has to overcome the resisting force from the vacuum. For the case at hand, one can observe that $b \sim 2$, which suggests that, as $|\alpha_+|$ is always much smaller than $1/4$, the wall can expand within the LTE approach.

Combining the results from the Collisionless and LTE approach, one can expect the wall to reach a steady state rather than run away.

\end{widetext}

\end{appendix}

\end{document}